\newif\ifcol
\coltrue

\newif\ifbw
\bwfalse

\newif\ifdraft
\draftfalse
\newcommand{\pagesize}{a4paper}
\ifdraft
  \colfalse
\fi

\newlength{\plotwidth}
\ifdraft
\fi

\ifdraft
  \documentclass[11pt,draftcls,\pagesize,onecolumn,twoside]{IEEEtran}
\else
  \documentclass[10pt,journal,\pagesize,twoside]{IEEEtran}
\fi

\usepackage{url}

\usepackage{ifpdf}

\ifpdf
  \usepackage[pdftex]{hyperref}
\else
  \usepackage[ps2pdf]{hyperref}
\fi

\usepackage{booktabs}

\usepackage{cite}      % Written by Donald Arseneau
\ifpdf
  \usepackage[pdftex]{graphicx}
\else
  \usepackage[dvips]{graphicx}
\fi

\usepackage{subfigure} % Written by Steven Douglas Cochran

\usepackage{amssymb}    % Required for some math symbols uses, e.g. R for real numbers.

\usepackage{amsmath}    % From the American Mathematical Society
\usepackage{bbold}
\usepackage{algpseudocode}

\newcommand{\eqn}[1]{(#1)}

\newcommand{\fig}[1]{Fig.~#1}

\newcommand{\sectn}[1]{Sec.~#1}

\newcommand{\etal}{\mbox{\it et al.}}
\newcommand{\eg}{\mbox{\it e.g.}}
\newcommand{\ie}{\mbox{\it i.e.}}

\newcommand{\snr}{{\rm SNR}}

\newcommand{\spcend}{\ensuremath{\:}}
\newcommand{\img}{\ensuremath{{\rm i}}}
\newcommand{\cconj}{\ensuremath{\ast}} 
\newcommand{\adjoint}{\ensuremath{{}^\dagger}} 
\newcommand{\reals}{\ensuremath{\mathbb{R}}}

\newcommand{\integers}{\ensuremath{\mathbb{Z}}}
\newcommand{\naturals}{\ensuremath{\mathbb{N}}}
\newcommand{\complex}{\ensuremath{\mathbb{C}}}
\newcommand{\ltwo}{\ensuremath{\mathrm{L}^2}}
\newcommand{\sphere}{\ensuremath{{\mathrm{S}^2}}}

\newcommand{\vect}[1]{\ensuremath{\mbox{\boldmath ${#1}$}}}

\newcommand{\dx}{\ensuremath{\mathrm{\,d}}}
\newcommand{\dmu}[1]{\ensuremath{\dx \Omega(#1)}}
\newcommand{\dmun}{\ensuremath{\dx \Omega}}

\newcommand{\innerp}[2]{\ensuremath{\langle {#1},\: {#2} \rangle}}

\newcommand{\saa}{\ensuremath{\theta}}
\newcommand{\sab}{\ensuremath{\varphi}}
\newcommand{\sas}{\ensuremath{\saa, \sab}}

\newcommand{\el}{\ensuremath{\ell}}
\newcommand{\m}{\ensuremath{m}}
\newcommand{\n}{\ensuremath{n}}
\newcommand{\spin}{\ensuremath{s}}

\newcommand{\elmax}{\ensuremath{{L}}}
\newcommand{\p}{\ensuremath{^\prime}}

\newcommand{\kron}[2]{\ensuremath{\delta_{{#1}{#2}}}}

\renewcommand{\exp}[1]{\ensuremath{{\rm e}^{#1}}}
\newcommand{\shfarg}[3]{\ensuremath{Y_{#1#2}({#3})}}
\newcommand{\shfargc}[3]{\ensuremath{Y_{#1#2}^\cconj({#3})}}

\newcommand{\shf}[2]{\ensuremath{Y_{#1#2}}}

\newcommand{\shc}[3]{\ensuremath{{#1}_{{#2}{#3}}}}
\newcommand{\shcc}[3]{\ensuremath{{#1}_{{#2}{#3}}^\cconj}}
\newcommand{\shcsp}[3]{\ensuremath{{#1}_{{#2},{#3}}}}

\newcommand{\dmatsmall}{\ensuremath{d}}
\newcommand{\dlmn}{\ensuremath{ \dmatsmall_{\m\n}^{\el} }}

\newcommand{\dlmnhalfpi}[3]{\ensuremath{ \Delta_{{#2}{#3}}^{#1} }}

\newcommand{\dlmnhalfpim}{\ensuremath{ \Delta_{{\m\p}{\m}}^{\el} }}
\newcommand{\dlmnhalfpisn}{\ensuremath{ \Delta_{{\m\p}{,-\spin}}^{\el} }}

\newcommand{\f}{\ensuremath{f}}
\newcommand{\fs}{\ensuremath{{}_\spin f}}

\newcommand{\nl}{\ensuremath{\sqrt{\frac{2\el+1}{4\pi}}}}

\newcommand{\G}[3]{\ensuremath{{{}_{#1} G_{{#2} {#3}}}}}

\newcommand{\Gsmm}{\ensuremath{\G{\spin}{\m}{\m\p}}}

\newcommand{\F}[3]{\ensuremath{{{}_{#1} F_{{#2} {#3}}}}}
\newcommand{\rF}[3]{\ensuremath{{{}_{#1} \tilde{F}_{{#2} {#3}}}}}
\newcommand{\Fsm}{\ensuremath{\F{\spin}{\m}{}}}
\newcommand{\rFsm}{\ensuremath{\rF{\spin}{\m}{}}}
\newcommand{\Fsmm}{\ensuremath{\F{\spin}{\m}{\m\p}}}

\newcommand{\Fsmmp}{\ensuremath{\F{\spin}{\m}{\m{\p}{\p}}}}

\newcommand{\saai}{\ensuremath{t}}
\newcommand{\sabi}{\ensuremath{p}}
\newcommand{\saaiang}{\ensuremath{\saa_\saai}}
\newcommand{\sabiang}{\ensuremath{\sab_\sabi}}

\newcommand{\saisang}{\ensuremath{\saaiang,\sabiang}}

\newcommand{\sumltrunc}{\ensuremath{\sum_{\el=0}^{\elmax-1}}}
\newcommand{\summtrunc}{\ensuremath{\sum_{\m=-(\elmax-1)}^{\elmax-1}}}
\newcommand{\summptrunc}{\ensuremath{\sum_{\m\p=-(\elmax-1)}^{\elmax-1}}}
\newcommand{\qweight}{\ensuremath{q}}

\newcommand{\N}{\ensuremath{{N}}}
\newcommand{\Ngl}{\ensuremath{{N_{\rm GL}}}}
\newcommand{\Ndh}{\ensuremath{{N_{\rm DH}}}}

\newcommand{\Nmw}{\ensuremath{{N_{\rm MW}}}}

\newcommand{\weight}{\ensuremath{w}}
\newcommand{\order}{\ensuremath{\mathcal{O}}}

\newcommand{\nmeas}{\ensuremath{M}}

\newcommand{\sparsity}{\ensuremath{K}}

\newcommand{\sensmat}{\ensuremath{\Phi}}

\newcommand{\opshtinv}{\ensuremath{\Lambda}}
\newcommand{\opshtfwd}{\ensuremath{\Gamma}}
\newcommand{\opshtcse}{\ensuremath{\Pi}}

\renewcommand{\shc}[3]{\ensuremath{{\hat{#1}}_{{#2}{#3}}}}
\renewcommand{\shcc}[3]{\ensuremath{{\hat{#1}}_{{#2}{#3}}^\cconj}}
\renewcommand{\shcsp}[3]{\ensuremath{{\hat{#1}}_{{#2},{#3}}}}

\renewcommand{\sectn}[1]{Section~#1}

\renewcommand{\fig}[1]{Figure~#1}

\usepackage{color}

\begin{document}
\title{Sparse image reconstruction on the sphere: implications of a new
  sampling theorem}
%
%
% author names and IEEE memberships
% note positions of commas and nonbreaking spaces ( ~ ) LaTeX will not break
% a structure at a ~ so this keeps an author's name from being broken across
% two lines.
% use \thanks{} to gain access to the first footnote area
% a separate \thanks must be used for each paragraph as LaTeX2e's \thanks
% was not built to handle multiple paragraphs

% 

\author{Jason~D.~McEwen,~\IEEEmembership{Member,~IEEE,} 
  Gilles~Puy,
  Jean-Philippe Thiran,~\IEEEmembership{Senior Member,~IEEE,}\\
  Pierre Vandergheynst,
  Dimitri Van De Ville,~\IEEEmembership{Senior Member,~IEEE,}
  and Yves~Wiaux,~\IEEEmembership{Member,~IEEE}%
%        Michael~Shell,~\IEEEmembership{Member,~IEEE,}
%        John~Doe,~\IEEEmembership{Fellow,~OSA,}
%        and~Jane~Doe,~\IEEEmembership{Life~Fellow,~IEEE}% <-this % stops a space
  \thanks{Manuscript received August 7, 2012; revised December 20,
    2012; accepted February 15, 2013.  Date of publication February 26
    2013; date of current version April 12, 2013.  We gratefully
    acknowledge use of some of the data and software made available on
    Frederik Simons' webpage: {http://www.frederik.net}. The work of
    J.~D.~McEwen is supported by the Swiss National Science Foundation
    (SNSF) under grant 200021-130359 and by a Newton International
    Fellowship from the Royal Society and the British Academy.  The
    work of Y.~Wiaux is supported in part by the Center for Biomedical
    Imaging (CIBM) of the Geneva and Lausanne Universities, Ecole
    Polytechnique F{\'e}d{\'e}rale de Lausanne (EPFL), and the
    Leenaards and Louis-Jeantet foundations, and in part by the SNSF
    under grant PP00P2-123438.}%
  \thanks{Copyright (c) 2013 IEEE. Personal use of this material is
    permitted. However, permission to use this material for any other
    purposes must be obtained from the IEEE by sending a request to
    pubs-permissions@ieee.org.}%
  \thanks{J.~D.~McEwen is with the Department of Physics and
    Astronomy, University College London, London WC1E 6BT, UK. G.~Puy,
    J.-Ph.~Thiran, P.~Vandergheynst and Y.~Wiaux are with the Signal
    Processing Laboratories, Institute of Electrical Engineering,
    Ecole Polytechnique F{\'e}d{\'e}rale de Lausanne (EPFL), CH-1015
    Lausanne, Switzerland.  D.~Van De Ville is with the Institute of
    Bioengineering, EPFL, CH-1015 Lausanne, Switzerland, and with the
    Department of Radiology and Medical Informatics, University of
    Geneva (UniGE), CH-1211 Geneva, Switzerland.  G.~Puy is also with
    the Institute of the Physics of Biological Systems, EPFL, CH-1015
    Lausanne, Switzerland. Y.~Wiaux is also with the Institute of
    Bioengineering, EPFL, CH-1015 Lausanne, Switzerland, and with the
    Department of Radiology and Medical Informatics, University of
    UniGE, CH-1211 Geneva, Switzerland.}%
  \thanks{E-mail: jason.mcewen@ucl.ac.uk (J.~D.~McEwen)}}
% note the % following the last \IEEEmembership and also the first \thanks - 
% these prevent an unwanted space from occurring between the last author name
% and the end of the author line. i.e., if you had this:
% 
% \author{....lastname \thanks{...} \thanks{...} }
%                     ^------------^------------^----Do not want these spaces!
%
% a space would be appended to the last name and could cause every name on that
% line to be shifted left slightly. This is one of those "LaTeX things". For
% instance, "A\textbf{} \textbf{}B" will typeset as "A B" not "AB". If you want
% "AB" then you have to do: "A\textbf{}\textbf{}B"
% \thanks is no different in this regard, so shield the last } of each \thanks
% that ends a line with a % and do not let a space in before the next \thanks.
% Spaces after \IEEEmembership other than the last one are OK (and needed) as
% you are supposed to have spaces between the names. For what it is worth,
% this is a minor point as most people would not even notice if the said evil
% space somehow managed to creep in.
%
% The paper headers
\markboth{IEEE Transactions on Image Processing,~Vol.~22, No.~6,~June~2013}%
{McEwen \etal: Sparse image reconstruction on the sphere}
% The only time the second header will appear is for the odd numbered pages
% after the title page when using the twoside option.
% 
% *** Note that you probably will NOT want to include the author's name in ***
% *** the headers of peer review papers.                                   ***

% If you want to put a publisher's ID mark on the page
% (can leave text blank if you just want to see how the
% text height on the first page will be reduced by IEEE)
%\pubid{0000--0000/00\$00.00~\copyright~2002 IEEE}

% use only for invited papers
%\specialpapernotice{(Invited Paper)}

% make the title area
\maketitle

%------------------------------------------------------------------------------

% Abstract.
%------------------------------------------------------------------------------
% Abstract
%------------------------------------------------------------------------------

\begin{abstract}  
  We study the impact of sampling theorems on the fidelity of sparse
  image reconstruction on the sphere.  We discuss how a reduction in
  the number of samples required to represent all information content
  of a band-limited signal acts to improve the fidelity of sparse
  image reconstruction, through both the dimensionality and sparsity
  of signals.  To demonstrate this result we consider a simple
  inpainting problem on the sphere and consider images sparse in the
  magnitude of their gradient.  We develop a framework for total
  variation (TV) inpainting on the sphere, including fast methods to
  render the inpainting problem computationally feasible at
  high-resolution.  Recently a new sampling theorem on the sphere was
  developed, reducing the required number of samples by a factor of
  two for equiangular sampling schemes.  Through numerical simulations
  we verify the enhanced fidelity of sparse image reconstruction due
  to the more efficient sampling of the sphere provided by the new
  sampling theorem.
\end{abstract}

% ctrl+c _
%%% Local Variables: 
%%% mode: latex
%%% TeX-master: "css2"
%%% End: 3

% Keywords.

% For peerreview papers, inserts a page break and creates the second title.
% Will be ignored for other modes.
\IEEEpeerreviewmaketitle

% Main body.

%==============================================================================
\section{Introduction}
%==============================================================================

\PARstart{I}{mages} are observed on a spherical manifold in many
fields, from astrophysics (\eg\ \cite{jaroski:2010}) and biomedical
imaging (\eg\ \cite{johansenberg:2009}), to computer graphics (\eg\
\cite{ramamoorthi:2004}) and beyond.  In many of these settings
inverse problems arise, where one seeks to recover an unknown image
from linear measurements, which may be noisy, incomplete or acquired
through a convolution process, for example.  Such inverse problems are
typically solved by assuming some regularising prior on the unknown
image to be recovered.

Sparsity priors have received a lot of attention recently, since (a)
they have been shown to be an effective and versatile approach for
representing many real-world signals and (b) a sound theoretical
foundation is provided by the emerging and rapidly evolving theory of
compressive sensing \cite{candes:2006a,candes:2006c,donoho:2006}.  On
the sphere, compressive sensing has been considered for signals sparse
in the spherical harmonic domain \cite{rauhut:2011}, however a general
theoretical framework does not yet exist for signals sparse in
spatially localised representations.  Nevertheless, sparse image
reconstruction on the sphere in alternative representations, such as a
set of overcomplete dictionaries, may still be considered; indeed,
such an approach has been shown to be very effective
\cite{abrial:2007}.

Although compressive sensing goes beyond Nyquist sampling, the Nyquist
limit nevertheless defines the benchmark from which compressive
sensing improvements are relative.  Compressive sensing results are
thus tightly coupled to the underlying sampling theorem on the
manifold of interest.  On the sphere, unlike Euclidean space, the
number of samples required in the harmonic and spatial domains differ,
with different sampling theorems on the sphere requiring a different
number of samples in the spatial domain.  Consequently, the sampling
theorem adopted influences the performance of sparse signal
reconstruction on the sphere.  Studying the impact of
sampling theorems on the sphere on the performance of sparse signal
reconstruction is the focus of the current article.

When considering signal priors that incorporate spatially localised
information (for example directly in real space, in the magnitude of the
gradient of signals, or through a wavelet basis or overcomplete
dictionary), the sampling theorem that is adopted
becomes increasingly important. 
Recently, a new sampling theorem on the sphere was developed by two of
the authors of the current article for equiangular sampling schemes
\cite{mcewen:fssht}, reducing Nyquist sampling on the sphere by a
factor of two compared to the canonical approach
\cite{driscoll:1994,healy:2003}.  The reduction in the number of
samples required to represent a band-limited signal on the sphere has
important implications for sparse image reconstruction.

To gain some intuition regarding these implications, we appeal to
standard compressive sensing results in Euclidean space, where the
ratio of the number of measurements $\nmeas$ required to reconstruct a
sparse image, to its dimensionality $\N$, goes as $\nmeas/\N \propto
\sparsity$ \cite{candes:2006c,candes:2010}, where $\sparsity$ is the sparsity
measure of the image (\ie\ the number of non-zero coefficients in some
sparse representation).\footnote{Typically the mutual coherence
    of the measurement and sparsifying operators also plays a role
    \cite{candes:2006c}.  However, in Euclidean space, as on the
    sphere, discrete inner products can be related to the (unique)
    continuous inner product (via a sampling theorem).  Consequently,
    the measure of coherence is invariant to the choice of sampling
    theorem (the coherence is defined through the continuous inner
    product, which is the same for all sampling theorems).  For the
    purpose of comparing sampling theorems on the sphere, we can thus
    safely neglect the impact of coherence.}  If one is not concerned
with the number of measurements required to achieve a given reconstruction
fidelity but rather with the best 
fidelity for a given number of measurements, then this suggests
  reconstruction fidelity improves with decreasing dimensionality of
  the signal $\N$ and with decreasing sparsity $\sparsity$.

Both of these quantities, dimensionality and sparsity, are related to
the number of samples required to capture all information content of
the underlying signal, as prescribed by the adopted sampling
theorem.  Spatial dimensionality is given identically by the number of
samples of the sampling theorem.  For any sparse representation of an
image that captures spatially localised information, the sparsity of
the signal is also directly related to spatial sampling.  For example,
in a wavelet representation, wavelets are located on each sample
point. A less dense dictionary of wavelet atoms required to span
  the space will lead to a more sparse representation of images when
  the sparsity is computed in an analysis approach, \ie\ as the number of
  non-zero projections of the signal onto the wavelet atoms.  We
  concentrate on such analysis priors here, as suggested by the recent
  evolution of compressive sensing with redundant dictionaries
  \cite{candes:2010}.  This argument can be extended to sparsity in
the gradient and, in fact, all sparsity measures that capture
spatially localised signal content.
Consequently, for images sparse in a spatially localised
representation, the ability to represent a band-limited signal on the
sphere with fewer samples while still capturing all of its information
content will improve the fidelity of sparse image reconstruction by
enhancing both the dimensionality and sparsity of signals.

In this article we study the implications of a new sampling theorem
\cite{mcewen:fssht} for sparse image reconstruction on the sphere.  We
verify the hypothesis that a more efficient sampling of the sphere, as
afforded by the new sampling theorem \cite{mcewen:fssht}, enhances the
fidelity of sparse image reconstruction through both the
dimensionality and sparsity of signals.  To demonstrate this result we
consider a simple inpainting problem, where we recover an image on the
sphere from incomplete spatial measurements.  We consider images
  sparse in the magnitude of their gradient, as an illustration of the
  general setting, and develop a framework for total variation (TV)
  inpainting on the sphere.  Solving these problems is
computationally challenging; hence we develop fast methods for this
purpose.  Our framework is general and is trivially extended to other
sparsity priors that incorporate spatially localised information.

The remainder of the article is structured as follows.  In
\sectn{\ref{sec:sampling}} we concisely review the harmonic structure
of the sphere and corresponding sampling theorems.  We develop a
framework for TV inpainting on the sphere in
\sectn{\ref{sec:compressive}}.  In \sectn{\ref{sec:algorithms}} we
describe algorithms for solving the optimisation problems on the
  sphere that arise in our TV inpainting framework. Numerical
simulations are performed in \sectn{\ref{sec:simulations}}, showing
the enhanced fidelity of sparse image reconstruction provided by a
more efficient sampling of the sphere.  Concluding remarks are made in
\sectn{\ref{sec:conclusions}}.

%==============================================================================
\section{Sampling on the Sphere}
\label{sec:sampling}
%==============================================================================

A sampling theorem on the sphere states that all information in a
(continuous) band-limited signal is captured in a finite number of
samples in the spatial domain.  Since a (continuous) band-limited
  signal on the sphere may be represented by a finite harmonic
  expansion, a sampling theorem on the sphere is equivalent to an
  exact prescription for computing a spherical harmonic transform from
  a finite set of spatial samples. In
this section we review the harmonic structure of the sphere, before
discussing sampling theorems on the sphere.

%==============================================================================
\subsection{Harmonic structure of the sphere}

We consider the space of square integrable functions on the sphere
$\ltwo(\sphere)$, with the inner product of $x,y\in\ltwo(\sphere)$
defined by
\begin{equation*}
\innerp{x}{y} \equiv \int_\sphere \dmu{\sas} \: x(\sas) \: y^\cconj(\sas) 
\spcend ,
\end{equation*}
where $\dmu{\sas} = \sin \saa \dx \saa \dx \sab$ is the usual
invariant measure on the sphere and $(\sas)$ denote spherical
coordinates with colatitude $\saa \in [0,\pi]$ and longitude $\sab \in
[0,2\pi)$.  Complex conjugation is denoted by the superscript
${}^\cconj$.
The canonical basis for the space of square integrable functions on
the sphere is given by the spherical harmonics $\shf{\el}{\m} \in
\ltwo(\sphere)$, with natural $\el\in\naturals$, integer
$\m\in\integers$ and $|\m|\leq\el$.  Due to the orthogonality and
completeness of the spherical harmonics, any square integrable
function on the sphere $x \in \ltwo(\sphere)$ may be represented by
its spherical harmonic expansion
\begin{equation}
\label{eqn:sht_inv}
x(\sas) = 
\sum_{\el=0}^\infty
\sum_{\m=-\el}^\el
\shc{x}{\el}{\m} \:
\shfarg{\el}{\m}{\sas}
\spcend ,
\end{equation}
where the spherical harmonic coefficients are given by the usual
projection onto each basis function: 
\begin{equation*}
\shc{x}{\el}{\m}
= \innerp{x}{\shf{\el}{\m}}
= \int_\sphere \dmu{\sas} \: x(\sas) \: \shfargc{\el}{\m}{\sas}
\spcend.  
\end{equation*}
Throughout, we consider signals on the sphere band-limited at
$\elmax$, that is signals such that $\shc{x}{\el}{\m}=0$, $\forall
\el\geq\elmax$, in which case the summation over \el\ in
\eqn{\ref{eqn:sht_inv}} may be truncated to the first \elmax\ terms.
Finally, note that the harmonic coefficients of a real function on the
sphere satisfy the conjugate symmetry relation
\mbox{$\shcc{x}{\el}{\m} = (-1)^\m \: \shcsp{x}{\el}{-\m}$}, which
follows directly from the conjugate symmetry of the spherical
harmonics.

%==============================================================================
\subsection{Sampling theorems on the sphere}

Sampling theorems on the sphere describe how to sample a band-limited
signal $x$ so that all information is contained in a finite number of
samples \N. Moreover, a sampling theorem on the sphere
  effectively encodes an exact quadrature rule for the integration of
  band-limited functions \cite{driscoll:1994,mcewen:fssht}. We denote
the concatenated vector of \N\ spatial measurements by
$\vect{x}\in\complex^{\N}$ and the concatenated vector of $\elmax^2$
harmonic coefficients by $\vect{\hat{x}}\in\complex^{\elmax^2}$.  The
number of spatial and harmonic elements, $\N$ and $\elmax^2$
respectively, may differ (and in fact do differ for all known sampling
theorems on the sphere).

Before discussing different sampling theorems on the sphere, we define
a generic notation to describe the harmonic transform corresponding to
a given sampling theorem.  A sampling theorem describes how to compute
the spherical harmonic transform of a signal exactly.  Since the
spherical harmonic transform and inverse are linear, we represent the
forward and inverse transform by the matrix operators
\mbox{$\opshtfwd\in\complex^{\elmax^2\times\N}$} and
$\opshtinv\in\complex^{\N\times\elmax^2}$ respectively.  The spherical
harmonic coefficients of a sampled signal (\ie\ image) on the sphere
$\vect{x}$ are given by the forward transform
\begin{equation*}
\vect{\hat{x}} = \opshtfwd \vect{x}
\spcend ,
\end{equation*}
while the original signal is recovered from its harmonic coefficients
by the inverse transform
\begin{equation*}
\vect{x} = \opshtinv \vect{\hat{x}}
\spcend .
\end{equation*}

Different sampling theorems then differ in the definition of
$\opshtinv$, $\opshtfwd$ and the number of spatial samples \N.  By
definition, all sampling theorems give exact spherical harmonic
transforms, implying $\opshtfwd \opshtinv = \mathbb{1}_{\elmax^2}$, where
$\mathbb{1}_k$ is the $k \times k$ identity matrix.  However, for all
sampling theorems on the sphere the number of samples required in the
spatial domain exceeds the number of coefficients in the harmonic
domain (\ie\ $\N>\elmax^2$), hence $\opshtinv \opshtfwd \neq
\mathbb{1}_{\N}$.  Consequently, for the \N\ sample positions of a
sampling theorem, an arbitrary set of sample values does not
necessarily define a band-limited signal (contrast this to the discrete
Euclidean setting where a finite set of samples uniquely defines a
band-limited signal).  Note also that the adjoint inverse (forward)
spherical harmonic transform differs to the forward (inverse)
spherical harmonic transform in the discrete setting.

For an equiangular sampling of the sphere, the Driscoll \& Healy (DH)
\cite{driscoll:1994} sampling theorem has become the standard,
requiring \mbox{$\Ndh = 2 \elmax (2 \elmax-1) \sim 4 \elmax^2$}
samples on the sphere to represent exactly a signal band-limited in
its spherical harmonic decomposition at \elmax.  Recently, a new
sampling theorem for equiangular sampling schemes has been developed
by McEwen \& Wiaux (MW) \cite{mcewen:fssht}, requiring only
\mbox{$\Nmw = (\elmax - 1) (2 \elmax - 1) + 1 \sim 2 \elmax^2$}
samples to represent a band-limited signal exactly.
No sampling theorem on the sphere reaches the optimal number of
samples suggested by the $\elmax^2$ dimension of a band-limited signal
in harmonic space (although the MW sampling theorem comes closest to
this bound).  The MW sampling theorem therefore achieves a more
efficient sampling of the sphere, with a reduction by a factor of
approximately two in the number of samples required to represent a
band-limited signal on the sphere.\footnote{Gauss-Legendre (GL)
quadrature can also be used to construct an efficient sampling theorem
on the sphere, with \mbox{$\Ngl = \elmax (2 \elmax-1) \sim 2
  \elmax^2$} samples (see \eg\ \cite{mcewen:fssht}).  The MW sampling
theorem nevertheless remains more efficient, especially at low
band-limits.  Furthermore, it is not so straightforward to define the
TV norm on the GL grid since it is not equiangular.  Finally,
algorithms implementing the GL sampling theorem have been shown to be
limited to lower band-limits and less accurate than the algorithms
implementing the MW sampling theorem \cite{mcewen:fssht}. Thus, we
focus on equiangular sampling theorems only in this article.}

Fast algorithms have been developed to compute forward and inverse
spherical harmonic transforms rapidly for both the DH
\cite{driscoll:1994,healy:2003} and MW \cite{mcewen:fssht} sampling
theorems.  These fast algorithms are implemented, respectively, in the
publicly available {\tt
  SpharmonicKit}\footnote{\url{http://www.cs.dartmouth.edu/~geelong/sphere/}}
package and the Spin Spherical Harmonic Transform ({\tt
  SSHT})\footnote{\url{http://www.jasonmcewen.org/}} package and are
essential to facilitate the application of these sampling theorems at
high band-limits.

%==============================================================================
\section{Sparse Image Reconstruction on the Sphere}
\label{sec:compressive}
%==============================================================================

A more efficient sampling of a band-limited signal on the sphere, as
afforded by the MW sampling theorem, improves the quality of sparse
image reconstruction for images that are sparse in a spatially
localised measure.  To demonstrate this result we consider a simple
inpainting problem on the sphere and consider images sparse in the
magnitude of their gradient.  We develop a framework for total
variation (TV) inpainting on the sphere, which relies on a sampling
theorem and its associated quadrature rule to define a discrete TV
norm on the sphere.  Firstly, we define the discrete TV norm on the
sphere, before secondly defining finite difference gradient operators
on the sphere.  Thirdly, we discuss the TV inpainting problem.

%==============================================================================
\subsection{TV norm on the sphere}
\label{sec:compressive:norm}

We define the discrete TV norm on the sphere by
\begin{equation}
\| \vect{x} \|_{\rm TV} 
\equiv
\sum_{\saai=0}^{\N_{\saa}-1} \:
\sum_{\sabi=0}^{\N_{\sab}-1} \:
\qweight(\saa_\saai) \:
| \nabla\vect{x} |
\spcend ,
\label{eqn:tvnorm_discrete}
\end{equation}
where $\saai$ and $\sabi$ index the equiangular samples in $\saa$ and
$\sab$ respectively, with the number of samples associated with a given
sampling theorem denoted in each dimension by
$\N_{\saa}$ and $\N_{\sab}$ respectively.
The discrete magnitude of the gradient is defined by
\begin{equation}
| \nabla\vect{x} |
\equiv
\sqrt{
\bigl ( \delta_{\saa} \vect{x} \bigr )^2
+
\frac{1}{\sin^2{\saa_\saai}}\bigl (  \delta_{\sab} \vect{x} \bigr )^2
} 
\spcend ,
\label{eqn:maggrad_discrete}
\end{equation}
to approximate the continuous magnitude of the gradient 
\begin{equation*}
| \nabla x | \equiv
\sqrt{
\Biggl ( \frac{\partial x}{\partial \saa} \Biggr)^2
+
\frac{1}{\sin^2\saa}\Biggl (  \frac{\partial x}{\partial \sab} \Biggr )^2
}
\end{equation*}
by finite differences.  The finite difference
operators $\delta_{\saa}$ and $\delta_{\sab}$ are defined explicitly
in the following subsection.
The contribution to the TV norm from the magnitude of the gradient for
each pixel value is weighted by the quadrature weights
$\qweight(\saa_\saai)$ of the sampling theorem adopted in order to
approximate continuous integration.\footnote{If the band-limiting
    operator $\Upsilon \equiv \opshtinv \opshtfwd
    \in\complex^{\N\times\N}$ were applied to $| \nabla\vect{x} |$ in
    \eqn{\ref{eqn:tvnorm_discrete}}, then the finite summation of
    \eqn{\ref{eqn:tvnorm_discrete}} would give an exact quadrature for
    the integral of the continuous function underlying the associated
    samples of the band-limited $| \nabla\vect{x} |$.  However,
    introducing the operator $\Upsilon$ in
    \eqn{\ref{eqn:tvnorm_discrete}} would make solving the
    optimisation problems defined subsequently problematic and would
    also prohibit passing the quadrature weights inside the gradient
    to eliminate numerical instabilities due to the $\sin\saa$ term.
    Consequently, we adopt the definition of the discrete TV norm on
    the sphere given by \eqn{\ref{eqn:tvnorm_discrete}}.  In any case,
    numerical experiments have shown that $\| \vect{x} \|_{\rm TV}$ is
    identical to $\sum_{\saai,\sabi} \qweight(\saa_\saai) \Upsilon |
    \nabla\vect{x} |$ to machine precision for the particular test
    images considered in \sectn{\ref{sec:simulations:lores}}.  Thus,
    the discrete TV norm defined by \eqn{\ref{eqn:tvnorm_discrete}}
    can be though of as an accurate proxy for $\int_\sphere \dmun \:
    \Upsilon | \nabla\vect{x} |$.}  The inclusion of the weights
$\qweight(\saa_\saai)$ also regularises the $\sin{\saa}$ term that
arises from the definition of the gradient on the sphere, eliminating
numerical instabilities that this would otherwise cause.

%==============================================================================
\subsection{Gradient operators on the sphere}
\label{sec:compressive:gradients}

The finite difference operators $\delta_{\saa}$ and $\delta_{\sab}$
defined on the sphere appear in the definition of the discrete
magnitude of the gradient given by \eqn{\ref{eqn:maggrad_discrete}},
and thus are required to compute the discrete TV norm on the sphere.
Furthermore, as we shall see, to solve the TV inpainting problems
outlined in the following subsection, the adjoints of these operators
are also required.  We define these operators and adjoints explicitly
here.

The operator $\delta_{\saa}$ is defined sample-wise by
\begin{align*}
 \vect{u}_{\saai,\sabi} 
 &\equiv
 ( \delta_{\saa} \vect{x} \bigr )_{\saai,\sabi} \\
 &\equiv
 \begin{cases}
   \: \vect{x}_{\saai+1,\sabi} - \vect{x}_{\saai,\sabi}, & \saai = 0,1,
   \cdots, \N_\saa-2 \mbox{ and } \forall \sabi\\
   \: 0, & \saai = \N_\saa-1 \mbox{ and } \forall \sabi
 \end{cases}
 \spcend ,
\end{align*}
with adjoint
\begin{align*}
 ( \delta_{\saa}^\dagger \vect{u} \bigr )_{\saai,\sabi}
 & =
 \begin{cases}
   \: -\vect{u}_{\saai,\sabi}, & \saai = 0 \mbox{ and } \forall \sabi\\
   \: \vect{u}_{\saai-1,\sabi} - \vect{u}_{\saai,\sabi}, & \saai = 1,\cdots\N_\saa-2 \mbox{ and } \forall \sabi\\
   \: \vect{u}_{\saai-1,\sabi}, & \saai = \N_\saa-1 \mbox{ and } \forall \sabi
 \end{cases}
 \spcend .
\end{align*}
Note that this definition is identical to the typical definition of
the corresponding operator on the plane \cite{chambolle:2004}.
The operator $\delta_{\sab}$ is defined sample-wise by
\begin{align*}
 \vect{v}_{\saai,\sabi} 
 &\equiv
 ( \delta_{\sab} \vect{x} \bigr )_{\saai,\sabi} \\
 &\equiv
 \begin{cases}
   \: \vect{x}_{\saai,\sabi+1} - \vect{x}_{\saai,\sabi}, & \sabi = 0,1,
   \cdots, \N_\sab-2 \mbox{ and } \forall \saai\\
   \: \vect{x}_{\saai,0} - \vect{x}_{\saai,\sabi}, & \sabi = \N_\sab-1 \mbox{ and } \forall \saai
 \end{cases}
 \spcend ,
\end{align*}
with adjoint
\begin{align*}
 ( \delta_{\sab}^\dagger \vect{v} \bigr )_{\saai,\sabi}
 & =
 \begin{cases}
   \: \vect{v}_{\saai,\N_\sab-1} - \vect{v}_{\saai,\sabi}, & \sabi = 0 \mbox{ and } \forall \saai\\
   \: \vect{v}_{\saai,\sabi-1} - \vect{v}_{\saai,\sabi}, & \sabi = 1,\cdots\N_\saa-1 \mbox{ and } \forall \saai
 \end{cases}
 \spcend .
\end{align*}
Since the sphere is periodic in \sab, we define the corresponding
finite difference operator to also be periodic.  The finite difference
operator and adjoint in \sab\ therefore differ to the typical
definition on the plane \cite{chambolle:2004}.

The TV norm on the sphere may then be seen as the sum of the magnitude
of the weighted gradient
\begin{equation*}
\| \vect{x} \|_{\rm TV} = 
\sum_{\saai=0}^{\N_{\saa}-1} \:
\sum_{\sabi=0}^{\N_{\sab}-1} \:
\bigl | \bigl ( \widetilde{\nabla} \vect{x} \bigr )_{\saai,\sabi} \bigr | 
\spcend ,
\end{equation*}
where 
\begin{equation*}
\bigl | \bigl ( \widetilde{\nabla} \vect{x} \bigr )_{\saai,\sabi} \bigr | 
= 
(\widetilde{\vect{u}}_{\saai,\sabi}^2 + \widetilde{\vect{v}}_{\saai,\sabi}^2)^{1/2}
\spcend ,
\end{equation*}
for 
\begin{equation*}
\begin{pmatrix}
 \widetilde{\vect{u}} \\
 \widetilde{\vect{v}} \\
\end{pmatrix} 
\equiv
\widetilde{\nabla} \vect{x} 
\spcend .
\end{equation*}
The weighted gradient operator is defined by 
\begin{equation*}
\widetilde{\nabla} \equiv
\begin{pmatrix}
 \widetilde{\delta}_{\saa} \\
 \widetilde{\delta}_{\sab} \\
\end{pmatrix} 
\spcend ,
\end{equation*}
where the weighted finite difference operators are defined by 
\begin{equation*}
\bigl (\widetilde{\delta}_{\saa} \bigr)_{\saai,\sabi} 
\equiv
\qweight(\saa_\saai) \bigl (\delta_{\saa} \bigr)_{\saai,\sabi} 
\end{equation*}
and
\begin{equation*}
\bigl (\widetilde{\delta}_{\sab}\bigr)_{\saai,\sabi} 
\equiv
\frac{\qweight(\saa_\saai)}{\sin{\saa_\saai}} \bigl (\delta_{\sab} \bigr)_{\saai,\sabi} 
\spcend .
\end{equation*}
Notice how the inclusion of the weights $\qweight(\saa_\saai)$
regularises the $\sin{\saa}$ term that arises from the definition of
the gradient on the sphere, eliminating numerical instabilities that
this would otherwise cause. If $\saa_\saai=\pi$, corresponding to the
South pole of the sphere, then $ ( \delta_{\sab} \vect{x} \bigr
)_{\saai,\sabi}=0$ and thus we define $ ( \widetilde{\delta}_{\sab}
\vect{x} \bigr )_{\saai,\sabi}=0$ to avoid dividing by
$\sin{\saa_\saai}=0$.  Note that the MW sampling theorem includes a
sample on the South pole, while the DH sampling theorem does not
(neither sampling theorem includes a sample on the North pole).
The adjoint weighted gradient operator is then applied as
\begin{equation*}
\vect{x}\p =
\widetilde{\nabla}^\dagger 
\begin{pmatrix}
 \widetilde{\vect{u}} \\
 \widetilde{\vect{v}} \\
\end{pmatrix} 
= 
\widetilde{\delta}_{\saa}^\dagger \widetilde{\vect{u}}
+
\widetilde{\delta}_{\sab}^\dagger \widetilde{\vect{v}}
\spcend ,
\end{equation*}
where the adjoint operators $\widetilde{\delta}_{\saa}^\dagger$ and
$\widetilde{\delta}_{\sab}^\dagger$ follow trivially from
$\delta_{\saa}^\dagger$ and $\delta_{\sab}^\dagger$.

%==============================================================================
\subsection{TV inpainting on the sphere}
\label{sec:compressive:inpainting}

We consider the measurement equation
\begin{equation*}
\vect{y} = 
\sensmat \vect{x} + \vect{n}
\spcend ,
\end{equation*}
where $\nmeas$ noisy real measurements $\vect{y}\in\reals^{\nmeas}$ of
the underlying real image on the sphere $\vect{x}\in\reals^{\N}$
are made.  The matrix implementing the measurement operator
$\sensmat\in\reals^{\nmeas \times \N}$ represents a uniformly random
masking of the image, with one non-zero, unit value on each row
specifying the location of the measured datum.  The noise
$\vect{n}\in\reals^\nmeas$ is assumed to be independent and
identically distributed (iid) Gaussian noise, with zero mean and
variance $\sigma_{\rm n}^2$.  We assume that the image $\vect{x}$ is
sparse in the norm of its gradient and thus attempt to recover
$\vect{x}$ from measurements $\vect{y}$ by solving the following TV
inpainting problem directly on the sphere:
\begin{equation}
\label{eqn:recon_spatial}
\vect{x}^\star =
\underset{\vect{x}}{\arg \min} \:
\| \vect{x} \|_{\rm TV} \:\: \mbox{such that} \:\:
\| \vect{y} - \sensmat \vect{x}\|_2 \leq \epsilon
\spcend .
\end{equation}
The square of the residual noise follows a scaled $\chi^2$
distribution with \nmeas\ degrees of freedom, \ie\ $\| \vect{y} -
\sensmat \vect{x}^\star\|_2^2 \sim \sigma_{\rm n}^2 \:\:
\chi^2(\nmeas)$.  Consequently, we choose $\epsilon^2$ to correspond
to the ($100\alpha$)th percentile of this distribution, giving a
probability $\alpha$ that pure noise produces a residual noise equal
to or smaller than the observed residual.  Note that the data
constraint in \eqn{\ref{eqn:recon_spatial}} is given by the usual
$\el_2$-norm, which is appropriate for Gaussian noise on a discrete
set of measurements.
Although we consider band-limited signals, we have not imposed this
constraint when solving \eqn{\ref{eqn:recon_spatial}}.  Consequently,
$\vect{x}^\star$ will not necessarily be band-limited at \elmax\ and
we impose this constraint on the solution by performing a forward and
inverse spherical harmonic transform: $\vect{x}^\star_{\elmax} =
\Upsilon \vect{x}^\star$, where the band-limiting operator
is defined by $\Upsilon \equiv \opshtinv \opshtfwd
\in\complex^{\N\times\N}$.

As discussed already, for images sparse in a measure that captures
spatially localised information, such as the TV norm, a more efficient
sampling of the signal enhances sparsity.  Furthermore, when
recovering signals in the spatial domain directly, the dimensionality
of the signal is also enhanced by a more efficient sampling.  These two
effects both act to improve the fidelity of sparse image
reconstruction.  Thus, the more efficient sampling of the MW sampling theorem
when compared to the DH sampling theorem will improve the fidelity of
sparse image reconstruction when solving the TV inpainting problem
given by \eqn{\ref{eqn:recon_spatial}}.  We verify these claims with
numerical experiments in \sectn{\ref{sec:simulations}}.

No sampling theorem on the sphere reaches the optimal number of
samples in the spatial domain suggested by the $\elmax^2$
dimensionality of the signal in the harmonic domain.  We may therefore
optimise the dimensionality of the signal that we attempt to recover
by recovering its harmonic coefficients $\hat{\vect{x}}$ directly.  We
do so by solving the following TV inpainting problem in harmonic
space:
\begin{equation}
\label{eqn:recon_harmonic}
\vect{\hat{x}}^{\prime\star} =
\underset{\vect{\hat{x}}^\prime}{\arg \min} \:
\| \opshtinv^\prime \vect{\hat{x}}^\prime \|_{\rm TV} \:\: \mbox{such that} \:\:
\| \vect{y} - \sensmat \opshtinv^\prime \vect{\hat{x}}^\prime\|_2 \leq \epsilon
\spcend .
\end{equation}
We impose reality of the recovered signal by explicitly imposing
conjugate symmetry in harmonic space through the conjugate symmetry
extension operator $\opshtcse \in \complex^{\elmax^2 \times
  \elmax(\elmax+1)/2}$, where $\opshtinv^\prime = \opshtinv \opshtcse$.  The
full set of harmonic coefficients of $x$ are given by
\mbox{$\vect{\hat{x}} = \opshtcse \vect{\hat{x}}^\prime$}, where
$\vect{\hat{x}}^\prime \in \complex^{\elmax(\elmax+1)/2}$ are the
harmonic coefficients for the spherical harmonic azimuthal index $\m$
non-negative only.  The image on the sphere is then recovered from
its harmonic coefficients by $\vect{x}^\star = \opshtinv^\prime
\vect{\hat{x}}^{\prime\star}$.  By solving the TV inpainting problem directly
in harmonic space, we naturally recover a signal band-limited at
\elmax.

When solving the TV inpainting problem \eqn{\ref{eqn:recon_harmonic}}
directly in harmonic space, the dimensionality of the recovered signal
is optimal and identical for both sampling theorems.  However, the
sparsity of the signal with respect to the TV norm remains enhanced
for the MW sampling theorem when compared to the DH sampling theorem.
Consequently, the MW sampling theorem will improve the fidelity of
sparse image reconstruction when solving the TV inpainting
problem given by \eqn{\ref{eqn:recon_harmonic}}, although through
sparsity only and not also dimensionality.  We verify these claims
with numerical experiments in \sectn{\ref{sec:simulations}}.  

Note that if a band-limit constraint were explicitly imposed in
problem \eqn{\ref{eqn:recon_spatial}}, then the two problems would be
equivalent, however, this would involve applying the band-limiting
operator $\Upsilon = \opshtinv \opshtfwd$, complicating the problem and increasing
the computational cost of finding a solution, while providing no
improvement over \eqn{\ref{eqn:recon_harmonic}}.  In the current
formulation of these two optimisation problems, problem
\eqn{\ref{eqn:recon_spatial}} has the advantage of simplicity, while
problem \eqn{\ref{eqn:recon_harmonic}} is the simplest formulation
that optimises dimensionality.

%==============================================================================
\section{Algorithms}
\label{sec:algorithms}
%==============================================================================

We solve the TV inpainting problems on the sphere given by
\eqn{\ref{eqn:recon_spatial}} and \eqn{\ref{eqn:recon_harmonic}} using
iterative convex optimisation methods. 
Solving the TV inpainting problem in harmonic space poses two
challenges as we go to high band-limits (\ie\ high-resolution).
Firstly, we require as an input to the convex optimisation algorithm
an upper bound on the inverse transform operator norm, which is
challenging to compute at high-resolution.  We describe a method to
compute the operator norm at high-resolution, which, crucially, does not
require an explicit computation of $\opshtinv$.  
Secondly, the inverse spherical harmonic transform $\opshtinv$ and its
adjoint operator $\opshtinv \adjoint$ must be applied repeatedly in
the iterative algorithm.  Fast algorithms are essential to perform
forward and inverse spherical harmonic transforms at high-resolution
and have been developed for both the DH
\cite{driscoll:1994,healy:2003} and MW \cite{mcewen:fssht} sampling
theorems.  To solve the inpainting problem at high-resolution we also
require a fast adjoint inverse transform.  We thus develop fast
algorithms to perform the adjoint forward and adjoint inverse
spherical harmonic transforms corresponding to the MW sampling
theorem.  Since we predict the MW sampling theorem to be superior to
the DH sampling theorem for sparse image reconstruction on the sphere
(a prediction that is validated by numerical experiments performed at
low band-limits (\ie\ low-resolution) in
\sectn{\ref{sec:simulations}}), we develop fast adjoint algorithms for
the MW sampling theorem only.
These methods then render the computation of solutions to the TV
inpainting problems feasible at high-resolution for the MW sampling
theorem.

%==============================================================================
\subsection{Convex optimisation}

We apply the Douglas-Rachford proximal splitting algorithm
\cite{combettes:2007} to solve the convex optimisation problems
(\ref{eqn:recon_spatial}) and
(\ref{eqn:recon_harmonic}).\footnote{We use Douglas-Rachford
    splitting since this does not require differentiability of the
    objective function and allows us to solve constrained optimisation
    problems where we adopt an indicator function to represent the
    measurement constraint.}  We describe only how to solve
  problem (\ref{eqn:recon_harmonic}) as problem
  (\ref{eqn:recon_spatial}) can be solved in the same way (by
  replacing $\opshtinv^\prime$ with the identity matrix
  $\mathbb{1}_{\N}$ and by replacing $\vect{\hat{x}}^\prime$ with
  $\vect{x}$).

The Douglas-Rachford algorithm \cite{combettes:2007} is based on a
splitting approach that requires the computation of two proximity
operators \cite{combettes:2011}. In our case, one proximity operator
is based on the TV norm $\| \opshtinv^\prime \, \vect{\cdot} \|_{\rm TV}$ and
the other on the data constraint $\| \vect{y} - \sensmat \opshtinv^\prime \,
\vect{\cdot}\|_2 \leq \epsilon$.

In the case of an image on the plane, the proximity operator based on
the TV norm may be computed using, for example, the method described
in \cite{chambolle:2004} or in \cite{beck:2009}.  For an image on the
sphere, the same methods can be used after introducing the following
modifications. 
In \cite{beck:2009} the algorithm to compute the proximity operator
of the TV norm is described in terms of a linear operator $\mathcal{L}$, its adjoint
$\mathcal{L}^\dagger$, and two projections onto a set $\mathcal{P}$
and a set $\mathcal{C}$. In our case, the linear operator
$\mathcal{L}$ and its adjoint $\mathcal{L}^\dagger$ may be redefined as
\begin{equation*}
\mathcal{L}: 
\begin{pmatrix}
 \widetilde{\vect{u}} \\
 \widetilde{\vect{v}} \\
\end{pmatrix}
 \longmapsto - \opshtinv^{\prime\dagger} \widetilde{\nabla}^{\dagger}
\begin{pmatrix}
 \widetilde{\vect{u}} \\
 \widetilde{\vect{v}} \\
\end{pmatrix}
\end{equation*}
and
\begin{equation*}
\mathcal{L}^\dagger: 
\vect{\hat{x}}^\prime \longmapsto 
- \widetilde{\nabla} \opshtinv^\prime
\vect{\hat{x}}^\prime
= -
\begin{pmatrix}
 \widetilde{\delta}_{\saa} \opshtinv^\prime \vect{\hat{x}}^\prime \\
 \widetilde{\delta}_{\sab} \opshtinv^\prime \vect{\hat{x}}^\prime \\
\end{pmatrix}
\spcend,
\end{equation*}
where the set $\mathcal{P}$ is the set of
weighted gradient-pairs $(\widetilde{\vect{u}}, \widetilde{\vect{v}})$
such that \mbox{$\widetilde{\vect{u}}^2_{\saai,\sabi} +
  \widetilde{\vect{v}}^2_{\saai,\sabi} \leq 1$} and $\mathcal{C}$ is
simply given by the space of the recovered vector $\vect{\hat{x}}$.

The second proximity operator, related to the data constraint $\|
\vect{y}~ - ~\sensmat \opshtinv^\prime \; \vect{\cdot}\|_2 \leq \epsilon$, is
computed using the method described in \cite{fadili:2009} directly.

%==============================================================================
\subsection{Operator norm bound}
\label{sec:algorithms:bound}

The convex optimisation algorithm requires as input upper bounds
for the norms of the operators that appear in the problem.  The
calculation of these norms is in most cases straightforward, however
the calculation of the inverse spherical harmonic transform operator norm, defined by
\begin{equation*}
\| \opshtinv \|_2 \equiv 
\max_{\| \vect{\hat{x}} \|_2 = 1} \| \opshtinv \vect{\hat{x}} \|_2
\spcend ,
\end{equation*}
can prove problematic. At low-resolution
$\| \opshtinv \|_2$ may be computed explicitly, however this is not
feasible at high-resolution since even
computing and storing \opshtinv\ explicitly is challenging.

We develop a method here to estimate this norm for the MW sampling
theorem without computing $\opshtinv$ explicitly.  We seek a sampled
function on the sphere $\vect{x} = \opshtinv \vect{\hat{x}}$ that
maximises $ \| \vect{x} \|_2$, while satisfying the constraint $\|
\vect{\hat{x}} \|_2 = 1$.  By the Parseval relation and the sampling
theorem on the sphere, this constraint may be rewritten:
\begin{eqnarray}
\| \vect{\hat{x}} \|_2 = 1 
&\underset{\mbox{Parseval}}{\Rightarrow} \nonumber &
\innerp{x}{x} = 1 \\
&\underset{\mbox{Sampling theorem}}{\Rightarrow} \nonumber &
\vect{x}_{\rm u}\adjoint Q_{\rm u} \vect{x}_{\rm u} = 1
\spcend ,
\end{eqnarray}
where $\vect{x}_{\rm u}\in\reals^{\N_{\rm u}}$ contains samples of
$x$, sampled at a resolution sufficient to represent $x^2$, \ie\
corresponding to band-limit $2\elmax-1$ (so that an exact quadrature
may be used to evaluate $\innerp{x}{x}$ from a discrete set of
samples), $Q_{\rm u} \in\reals^{\N_{\rm u} \times \N_{\rm u}}$ is the matrix
with corresponding quadrature weights along its diagonal, and where
$\N_{\rm u} \sim 2(2\elmax-1)^2 $.  Since we know that the quadrature
weights for the MW sampling theorem are closely approximated by
$\sin\saa$ \cite{mcewen:fssht}, the signal that maximises $ \|
\vect{x} \|_2$ while satisfying the constraint $\vect{x}_{\rm
  u}\adjoint Q_{\rm u} \vect{x}_{\rm u} = 1$ has its energy centred as much as
possible on the South pole since this is where the quadrature weights
are smallest (recall that the MW sampling scheme does not contain a
sample on the North pole).  This signal is given by the
  band-limited Dirac delta function centred on the South pole (see
  \eg\ \cite{simons:2006} for the definition of the band-limited Dirac
  delta function on the sphere).  The spherical harmonic coefficients
of this band-limited Dirac delta function $\delta^\elmax \in
\ltwo(\sphere)$ are given by
\begin{equation*}
\shc{\delta}{\el}{\m}^\elmax
= \kappa \: (-1)^\el \sqrt{\frac{2\el+1}{4\pi}} \kron{\m}{0}
\spcend ,
\end{equation*}
where $\kappa$ is a normalisation factor chosen to ensure $\|
\vect{\hat{\delta}}{}^\elmax \|_2 = 1$ and $\kron{i}{j}$ is the
Kronecker delta symbol.  The norm of the inverse spherical harmonic
transform operator may then be computed by $\| \opshtinv \|_2 \simeq \|
\opshtinv \vect{\hat{\delta}}{}^\elmax\|_2$, which, crucially, does not
require an explicit computation of $\opshtinv$, merely its application.

\begin{figure}
\centering
\includegraphics[width=85mm]{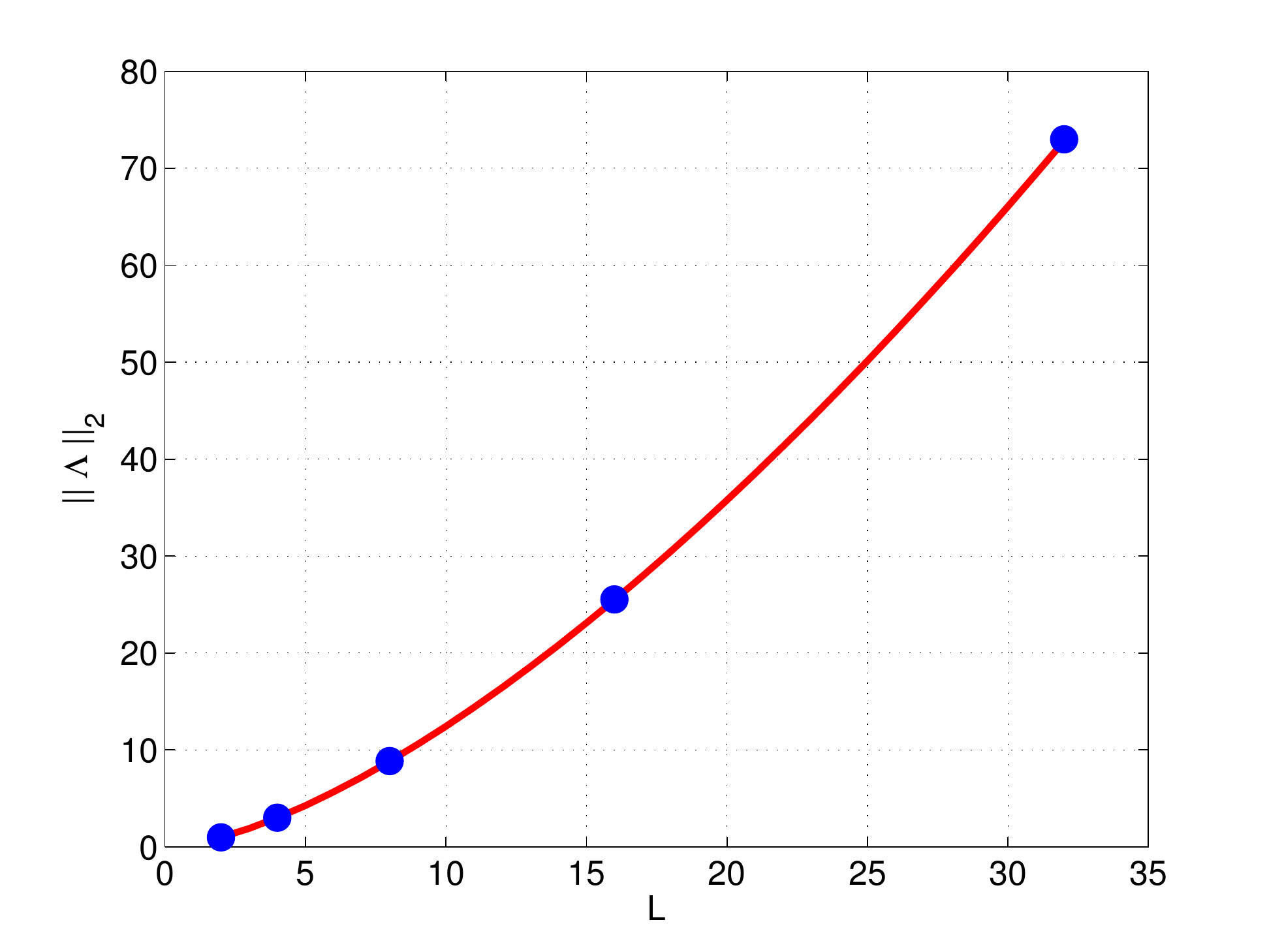}
\caption{Explicit calculation of the inverse spherical harmonic
  transform operator norm $\| \opshtinv \|_2$ and estimation by the
  method outlined in the text, at low-resolution.  The solid red line
  shows the estimated norm for all band-limits \elmax, while the solid
  blue circles show the values computed explicitly for
  \mbox{$\elmax\in\{2,4,8,16,32\}$}.  The estimated norm agrees with
  the actual norm very well.}
\label{fig:operator_norm}
\end{figure}

In \fig{\ref{fig:operator_norm}} we compute $\| \opshtinv \|_2$ by the
method outlined here and from $\opshtinv$ explicitly, for
low-resolution.  We find that the method to estimate the norm of the
inverse spherical harmonic transform operator outlined here estimates
the actual norm very well.  

We also derived an upper bound for the norm of this operator for the
MW sampling theorem.  However, the bound we derived is not tight and
we found empirically that the method outlined here to estimate the
norm itself, rather than a bound, is very accurate and improved the
performance of the optimisation algorithm considerably when compared
to a non-tight bound.  Although we do not prove so explicitly, we
conjecture that the method outlined here gives the inverse transform
operator norm exactly.

%==============================================================================
\subsection{Fast adjoint spherical harmonic transforms}
\label{sec:algorithms:adjoint}

Standard convex optimisation methods require not only the application
of the operators that appear in the optimisation problem but often
also their adjoints.  Moreover, these methods are typically iterative,
necessitating repeated application of each operator and its adjoint.
Thus, to solve optimisation problems that incorporate harmonic
transform operators, like the harmonic space TV inpainting problem
given by \eqn{\ref{eqn:recon_harmonic}}, fast algorithms to apply both
the operator and its adjoint are required to render high-resolution
problems computationally feasible.  

Here we develop fast algorithms to perform adjoint forward and adjoint
inverse spherical harmonic transforms for the MW sampling theorem.
Although we only require the adjoint inverse transform in this
article, for the sake of completeness we also derive a fast adjoint
forward transform.  Similarly, although we only consider scalar
functions in this article, for the sake of completeness we derive fast
adjoint algorithms for the spin setting. A spin function on the sphere
transforms as $\fs^\prime(\sas) = \exp{-\img \spin \chi} \: \fs(\sas)$
under a local rotation by $\chi \in [0,2\pi)$, where the prime denotes
the rotated function.  It is important to note that the rotation
considered here is \emph{not} a global rotation on the sphere but
rather a rotation by $\chi$ in the tangent plane at $(\sas)$ (see \eg\
\cite{mcewen:fssht} for further details).  In the expressions for the
fast algorithms derived below, the standard scalar case follows simply
by setting $\spin=0$.  These fast adjoint algorithms are implemented
in the publicly available {\tt
  SSHT}\footnote{\url{http://www.jasonmcewen.org/}} package
\cite{mcewen:fssht}.

The fast adjoint inverse spherical harmonic transform for the MW
sampling theorem follows by taking the adjoint of each stage of the
fast inverse transform \cite{mcewen:fssht} and applying these in
reverse order.  The final stage of the fast inverse transform
involves discarding out-of-domain samples and has adjoint
\begin{equation*}
{}_\spin \tilde{\f} \adjoint(\saisang) = 
 \begin{cases}
    \fs(\saisang) \: , & \saai \in \{ 0, 1, \dotsc, \elmax-1 \} \\
    0  \: , & \saai \in \{ \elmax, \dotsc, 2\elmax-2 \}
 \end{cases}
 \spcend .
\end{equation*}
The second stage of the fast adjoint inverse transform is given by
\begin{equation*}
  \Fsmm\adjoint = 
  \sum_{\saai=0}^{2\elmax-2} \:
  \sum_{\sabi=0}^{2\elmax-2} \:
  {}_\spin \tilde{\f} \adjoint(\saisang)  \:
  \exp{-\img (\m\p \saaiang + \m \sabiang)}
  \spcend ,
\end{equation*}
which may be computed rapidly using fast Fourier transforms (FFTs).  The
final stage of the fast adjoint inverse transform is given by
\begin{align*}
  {}_\spin \shc{\f}{\el}{\m}\adjoint \: =\: & (-1)^\spin \: \img^{\m+\spin} \nl \\
  &\times
  \summptrunc
  \dlmnhalfpim \:
  \dlmnhalfpisn \:
  \Fsmm\adjoint 
  \spcend ,
\end{align*}
where \mbox{$\dlmnhalfpi{\el}{\m}{\n} \equiv \dlmn (\pi/2)$} are the
Wigner $\dmatsmall$-functions evaluated for argument $\pi/2$ (see \eg\
\cite{varshalovich:1989}).  This final calculation dominates the
overall asymptotic complexity of the fast adjoint inverse transform,
resulting in an algorithm with complexity $\order(\elmax^3)$.

The fast adjoint forward spherical harmonic transform for the MW
sampling theorem follows by taking the adjoint of each stage of the
fast forward transform \cite{mcewen:fssht} and applying these in
reverse order.  The first stage of the fast adjoint forward transform
is given by
\begin{align*}
  \Gsmm\adjoint  \:=\: &(-1)^\spin \: \img^{-(\m+\spin)}  \\
  &\times \sumltrunc \nl \:
  \dlmnhalfpim \:
  \dlmnhalfpisn \:
   {}_\spin \shc{\f}{\el}{\m}
  \spcend .
\end{align*}
The next stage is given by the (reflected) convolution
\begin{equation*}
\Fsmmp\adjoint
 = 2 \pi \sum_{\m{\p}=-(\elmax-1)}^{\elmax-1} \: \Gsmm\adjoint \:
 \weight(\m\p - \m{\p}{\p})
\spcend ,
\end{equation*}
which is self-adjoint, followed by the
inverse Fourier transform in \saa\
\begin{equation*}
  \rFsm\adjoint(\saaiang) = 
  \frac{1}{2\elmax-1} \: \summptrunc \: \Fsmm\adjoint \: \exp{\img
    \m\p \saaiang}
  \spcend ,
\end{equation*}
which may be computed rapidly using FFTs.  The next stage consists of
the adjoint of the periodic extension of a function on the sphere
performed in the forward transform and is given by
\begin{align*}
 &\Fsm\adjoint(\saaiang)  = \\ 
 &\quad
 \begin{cases}
    \rFsm\adjoint(\saaiang) \\
    \quad + (-1)^{\m+\spin} \: \rFsm\adjoint(\saa_{2\elmax-2-\saai}) \: , & \saai \in \{ 0, 1, \dotsc, \elmax-2 \} \\
    \rFsm\adjoint(\saaiang) \: , & \saai = \elmax-1 
 \end{cases}
 \spcend .
\end{align*}
The final stage consists of the Fourier transform in \sab\
\begin{equation*}
  \fs\adjoint(\saisang) = 
  \frac{1}{2\elmax-1} \: \summtrunc \: 
  \Fsm\adjoint(\saaiang) \: \exp{\img \m \sabiang}
  \spcend ,
\end{equation*}
which may be computed rapidly using FFTs.  The first calculation
dominates the overall asymptotic complexity of the fast adjoint
forward transform, resulting in an algorithm with complexity
$\order(\elmax^3)$.

%==============================================================================
\section{Simulations}
\label{sec:simulations}
%==============================================================================

We perform numerical experiments to examine the impact of a more
efficient sampling of the sphere when solving the TV inpainting
problems defined in \sectn{\ref{sec:compressive}}.  Firstly, we perform a
low-resolution comparison of reconstruction fidelity when adopting the
DH and MW sampling theorems, where the predicted improvements in
reconstruction fidelity provided by the MW sampling theorem are verified
in practice.  Secondly, we perform a single simulation to illustrate TV
inpainting at high-resolution on a realistic test image.

%==============================================================================
\subsection{Low-resolution comparison on band-limited images}
\label{sec:simulations:lores}

\begin{figure}
\centering
\includegraphics[clip=,viewport=1 2 441 223,width=70mm]{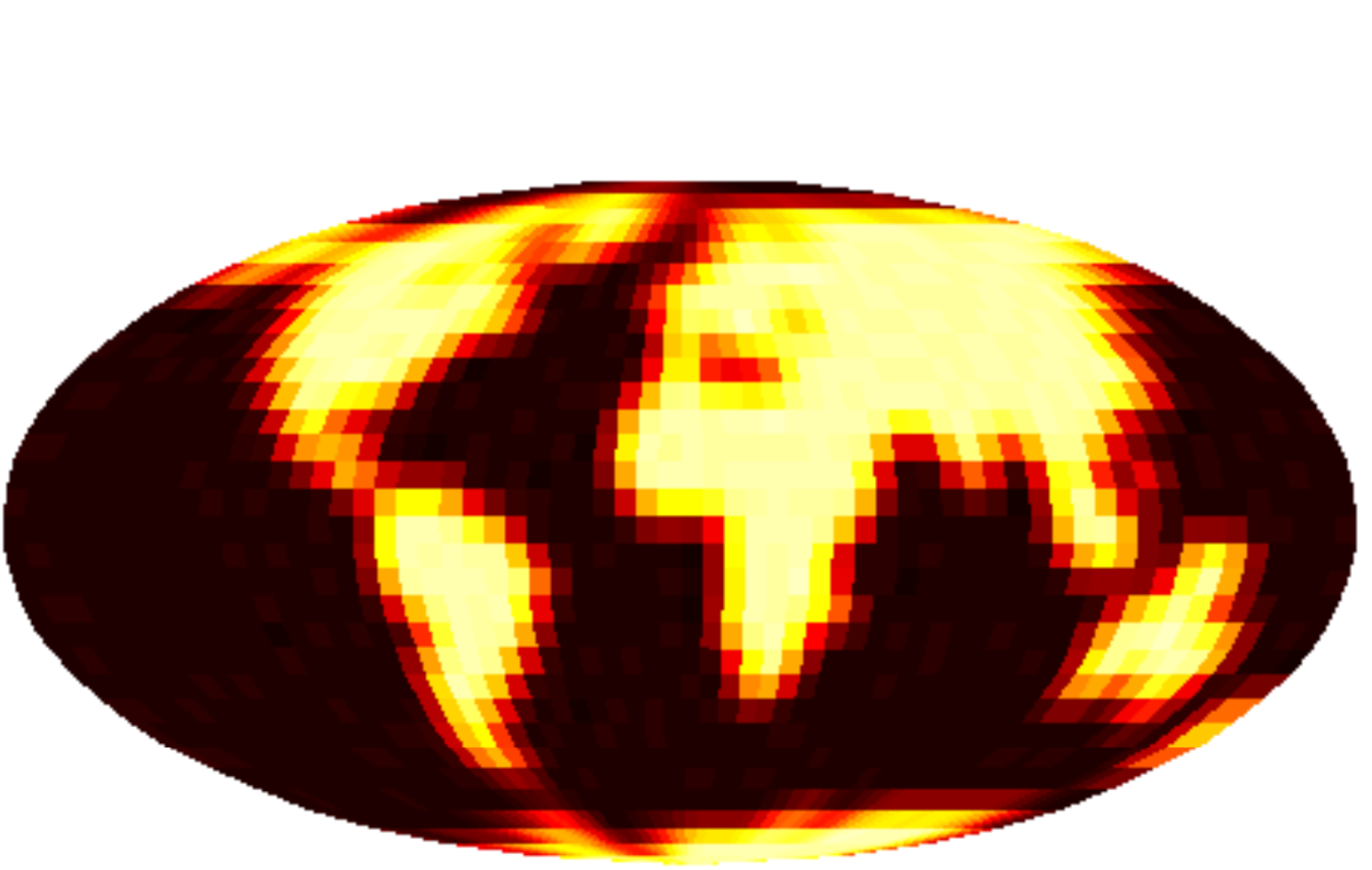}
\caption{Test image of Earth topographic data constructed to be
  sparse in its gradient and band-limited at $\elmax=32$.  This image
  constitutes the ground truth in our numerical experiments.  Here and
  subsequently data on the sphere are displayed using the Mollweide
  projection, with zero values shown in black, unit values shown in
  yellow, and the colour of intermediate values interpolated
  between these extremes.}
\label{fig:truth}
\end{figure}

A test image is constructed from Earth topography data.  The original
Earth topography data are taken from the Earth Gravitational Model
(EGM2008) publicly released by the U.S. National
Geospatial-Intelligence Agency (NGA) EGM Development
Team.\footnote{These data were downloaded and extracted using the
  tools available from Frederik Simons' webpage:
  \url{http://www.frederik.net}.}  To create a band-limited test
signal sparse in its gradient, the original data are thresholded at
their midpoint to create a binary Earth map (scaled to contain zero
and unit values), which is then smoothed by multiplication in harmonic
space with the Gaussian $\shc{G}{\el}{\m}={\rm exp}(-\el^2 \sigma_{\rm
  s})$, with $\sigma_{\rm s}=0.002$, to give a signal band-limited at
$\elmax=32$.  The resulting test image is displayed in
\fig{\ref{fig:truth}}. Let us stress that this test image is
constructed to satisfy the assumptions of our theoretical framework,
\ie\ the case of band-limited images that are sparse in their
gradient.  This is necessary to evaluate the theoretical predictions
based on our framework. A realistic test image is considered in the
following subsection.

Measurements of the test image are taken at uniformly random
locations on the sphere, as described by the measurement operator
$\sensmat$, in the presence of Gaussian iid noise with standard
deviation $\sigma_{\rm n}=0.01$.  Reconstructed images on the sphere
are recovered by solving the inpainting problems in the spatial and
harmonic domains, through \eqn{\ref{eqn:recon_spatial}} and
\eqn{\ref{eqn:recon_harmonic}} respectively, using both the DH and MW
sampling theorems, giving four reconstruction techniques.  The bound
$\epsilon$ is determined from $\alpha=0.99$.  We consider the
measurement ratios \mbox{$\nmeas/\elmax^2 \in \{1/4, \: 1/2,\: 1,\:
  3/2,\: \Nmw/\elmax^2 \sim 2 \}$} (recall that $\elmax^2$ is the
dimensionality of the signal in harmonic space).  The measurement
ratio $\nmeas/\elmax^2 = \Nmw/\elmax^2 \sim 2$ corresponds to complete
coverage for the MW sampling theorem, \ie\ Nyquist rate sampling on
the MW grid.

\begin{figure}
\centering
\includegraphics[width=95mm]{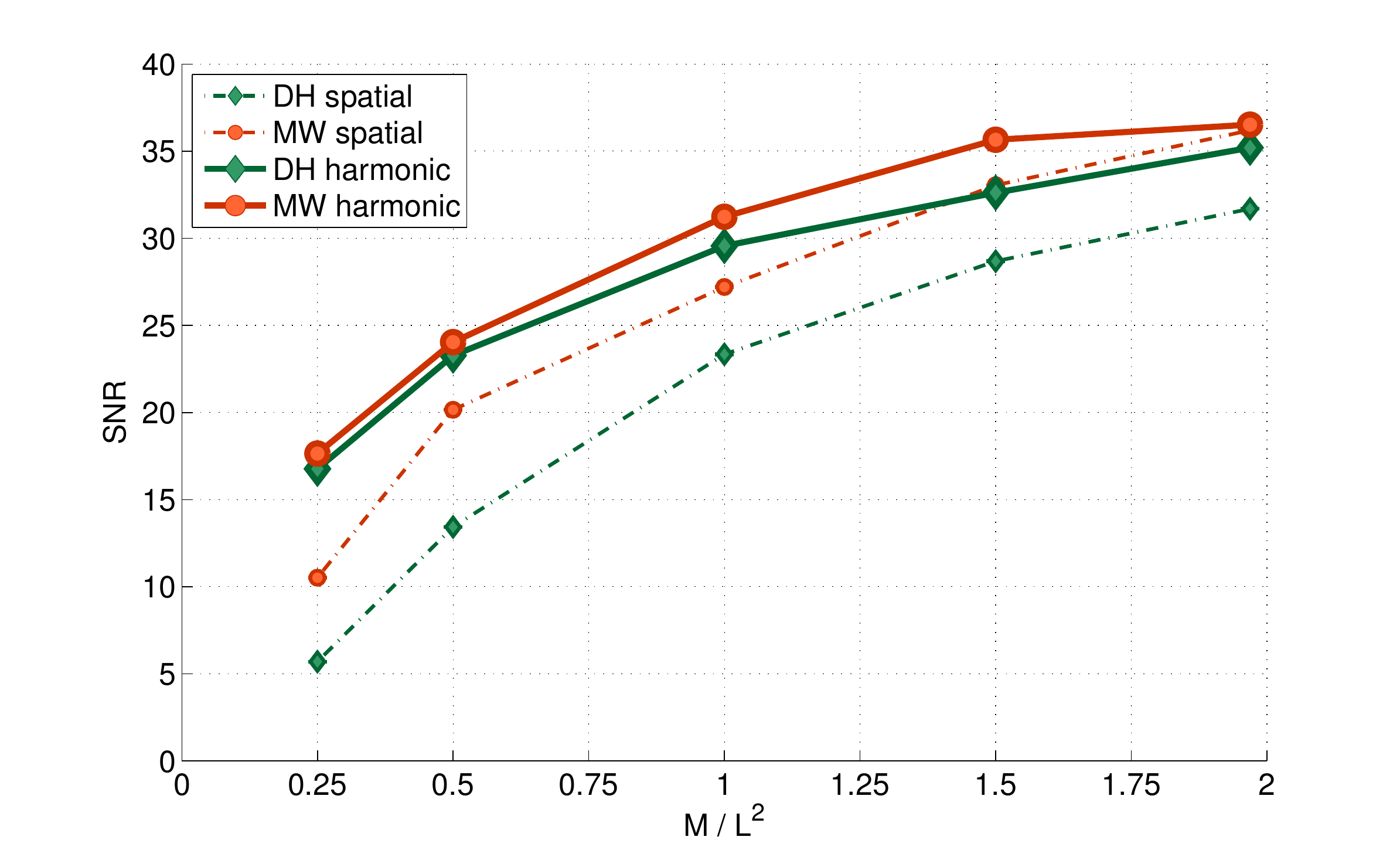}
\caption{Reconstruction performance for the DH (green/diamonds) and MW
  (red/circles) sampling theorems, when solving the TV inpainting
  problem in the spatial (dot-dashed line) and harmonic domain (solid
  line).  The MW sampling theorem provides enhancements in
  reconstruction quality when compared to the DH sampling theorem, due
  to dimensionality and sparsity improvements in spatial
  reconstructions, and due to sparsity (but not dimensionality)
  improvements in harmonic reconstructions.}
\label{fig:snr_vs_m}
\end{figure}

Typical reconstructed images are shown in
\fig{\ref{fig:reconstructions}} for the four reconstruction
techniques.
For each reconstruction technique and measurement ratio
$\nmeas/\elmax^2$, we perform ten simulations for random measurement
operators and noise.  To quantify the error of reconstruction, we
compute the signal-to-noise-ratio \mbox{$\snr=20 \log (\|
  \vect{\hat{x}} \|_2 / \| \vect{\hat{x}}^\star - \vect{\hat{x}}
  \|_2)$} (defined in harmonic space to avoid differences due to the
number of samples of each sampling theorem).  Note that the standard
$\el_2$-norm is used in the definition of the \snr\ given the discrete
nature of harmonic space on the sphere.  Reconstruction performance,
averaged over these ten simulations, is shown in
\fig{\ref{fig:snr_vs_m}}.

When solving the inpainting problem in the spatial domain through
\eqn{\ref{eqn:recon_spatial}} we see a large improvement in
reconstruction quality for the MW sampling theorem when compared to
the DH sampling theorem.  This is due to the enhancement in both
dimensionality and sparsity afforded by the MW sampling theorem in
this setting.  When solving the inpainting problem in the harmonic
domain through \eqn{\ref{eqn:recon_harmonic}} we see a considerable
improvement in reconstruction quality for each sampling theorem, since
we optimise the dimensionality of the recovered signal by going to
harmonic space.  For harmonic reconstructions, the MW sampling theorem
remains superior to the DH sampling theorem due to the enhancement in
sparsity (but not dimensionality) that it affords in this setting.
All of the predictions made in \sectn{\ref{sec:compressive}} are thus
exhibited in the numerical experiments performed in this section.  In
all cases, the superior performance of the MW sampling theorem is
clear.

\newlength{\sphereplotwidth}
\setlength{\sphereplotwidth}{41mm}

\begin{figure*}
\centering
\mbox{
\subfigure[DH spatial for $\nmeas/\elmax^2 = 1/4$]
  {
    \includegraphics[clip=,viewport=1 2 441 223,width=\sphereplotwidth]{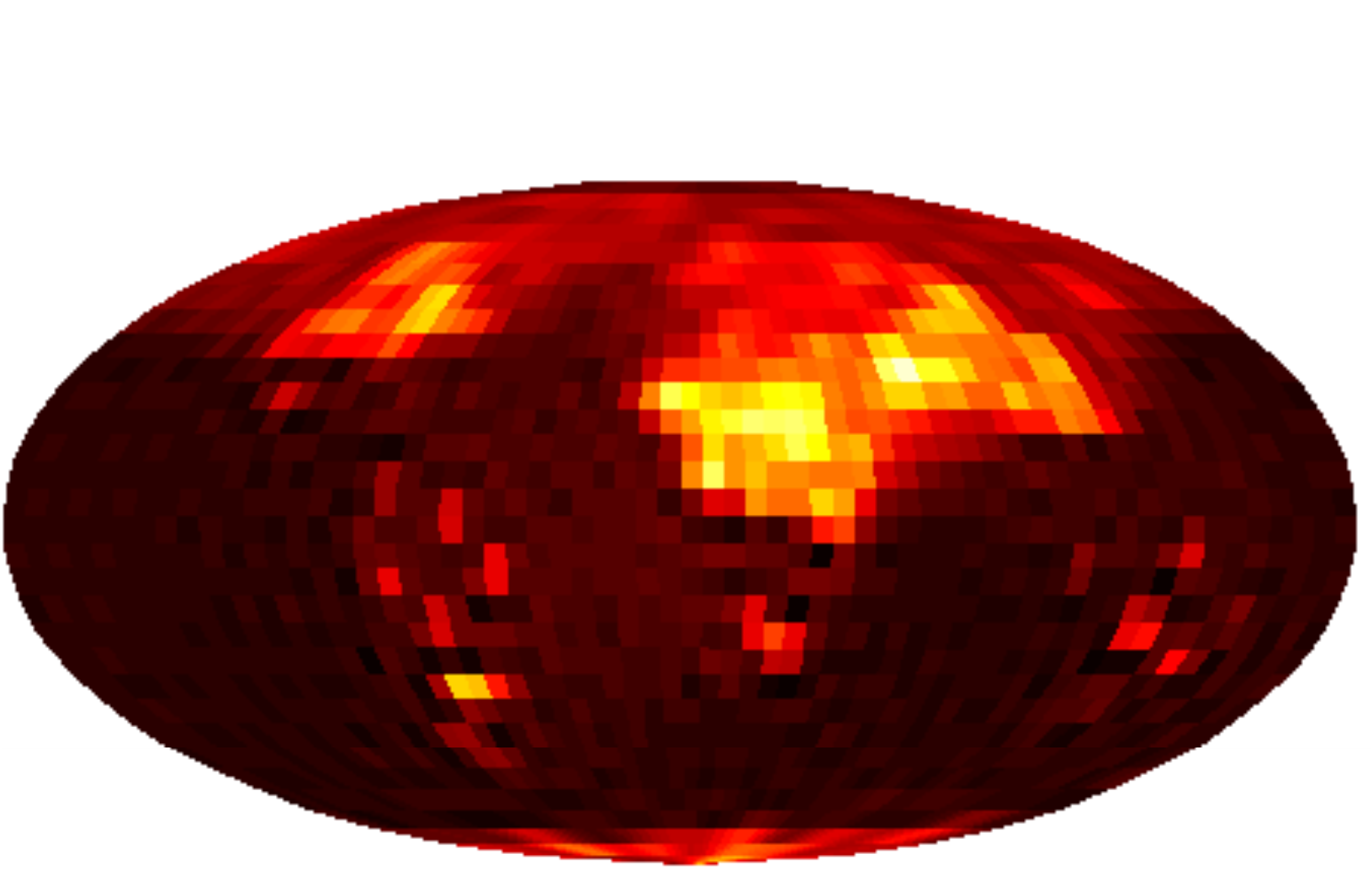}
  }
\subfigure[DH harmonic for $\nmeas/\elmax^2 = 1/4$]
  {
    \includegraphics[clip=,viewport=1 2 441 223,width=\sphereplotwidth]{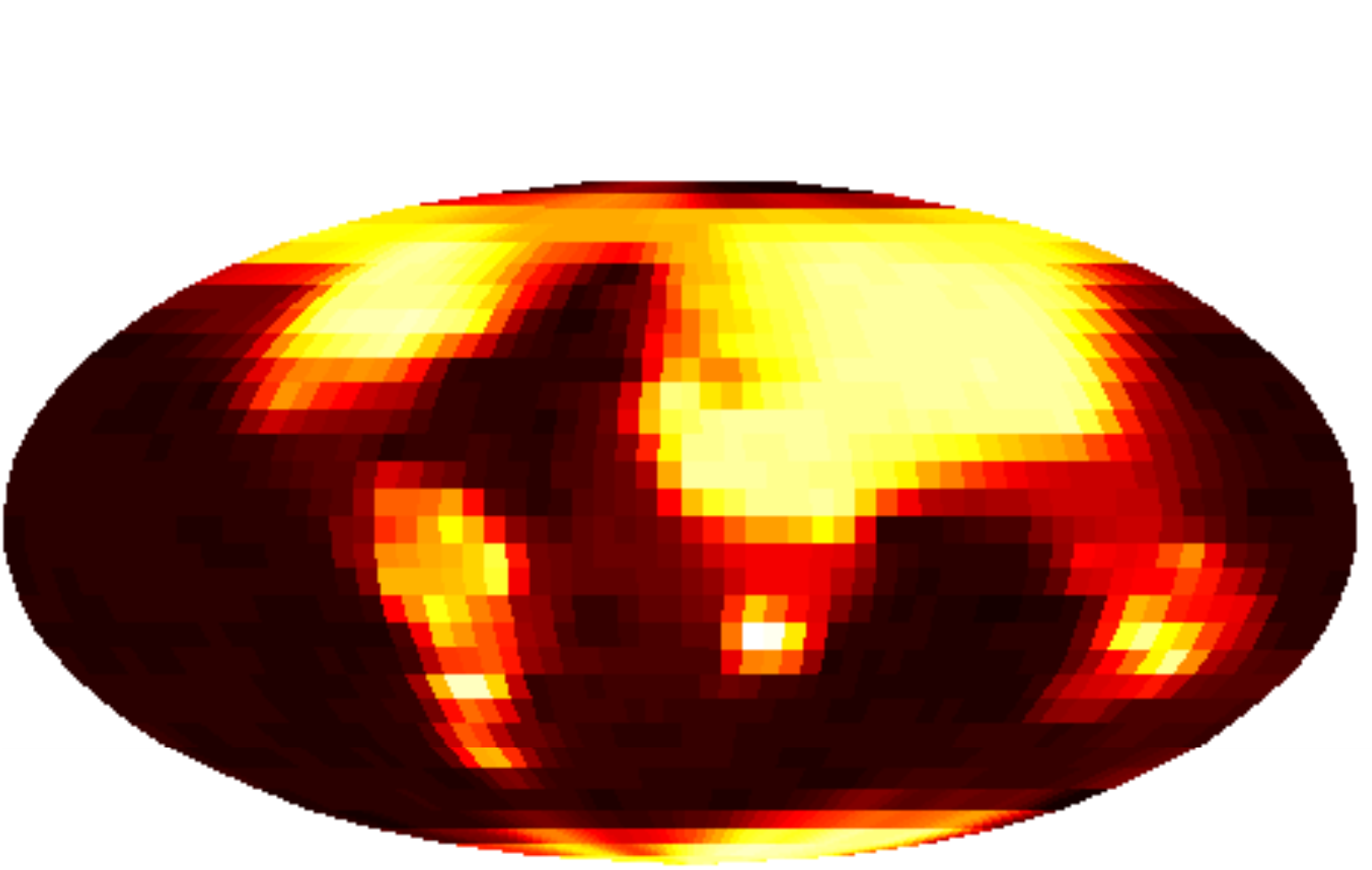}
  }
\subfigure[MW spatial for $\nmeas/\elmax^2 = 1/4$]
  {
    \includegraphics[clip=,viewport=1 2 441 223,width=\sphereplotwidth]{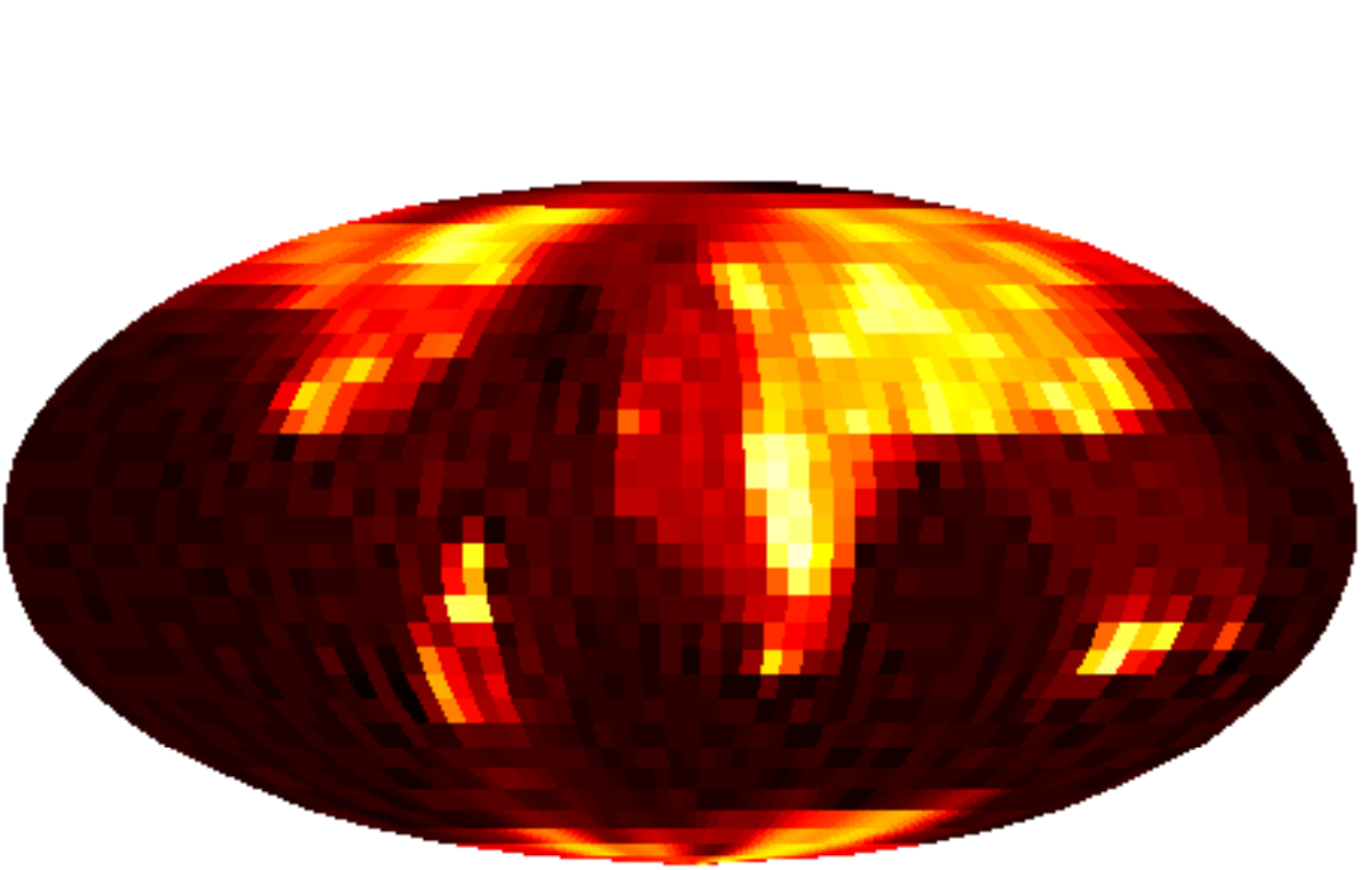}
  }
\subfigure[MW harmonic for $\nmeas/\elmax^2 = 1/4$]
  {
    \includegraphics[clip=,viewport=1 2 441 223,width=\sphereplotwidth]{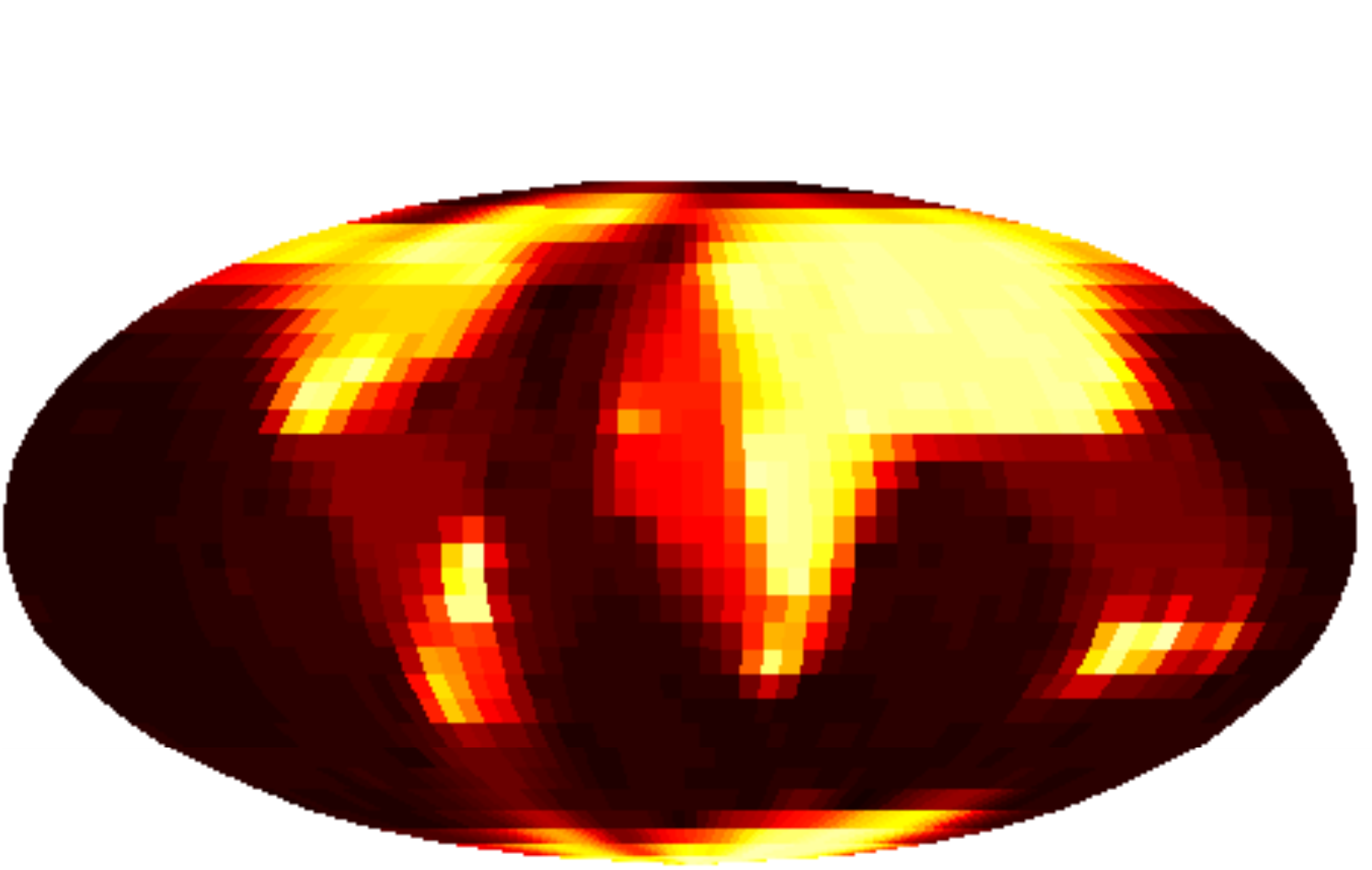}
  } 
}\\
\mbox{
\subfigure[DH spatial for $\nmeas/\elmax^2 = 1/2$]
  {
    \includegraphics[clip=,viewport=1 2 441 223,width=\sphereplotwidth]{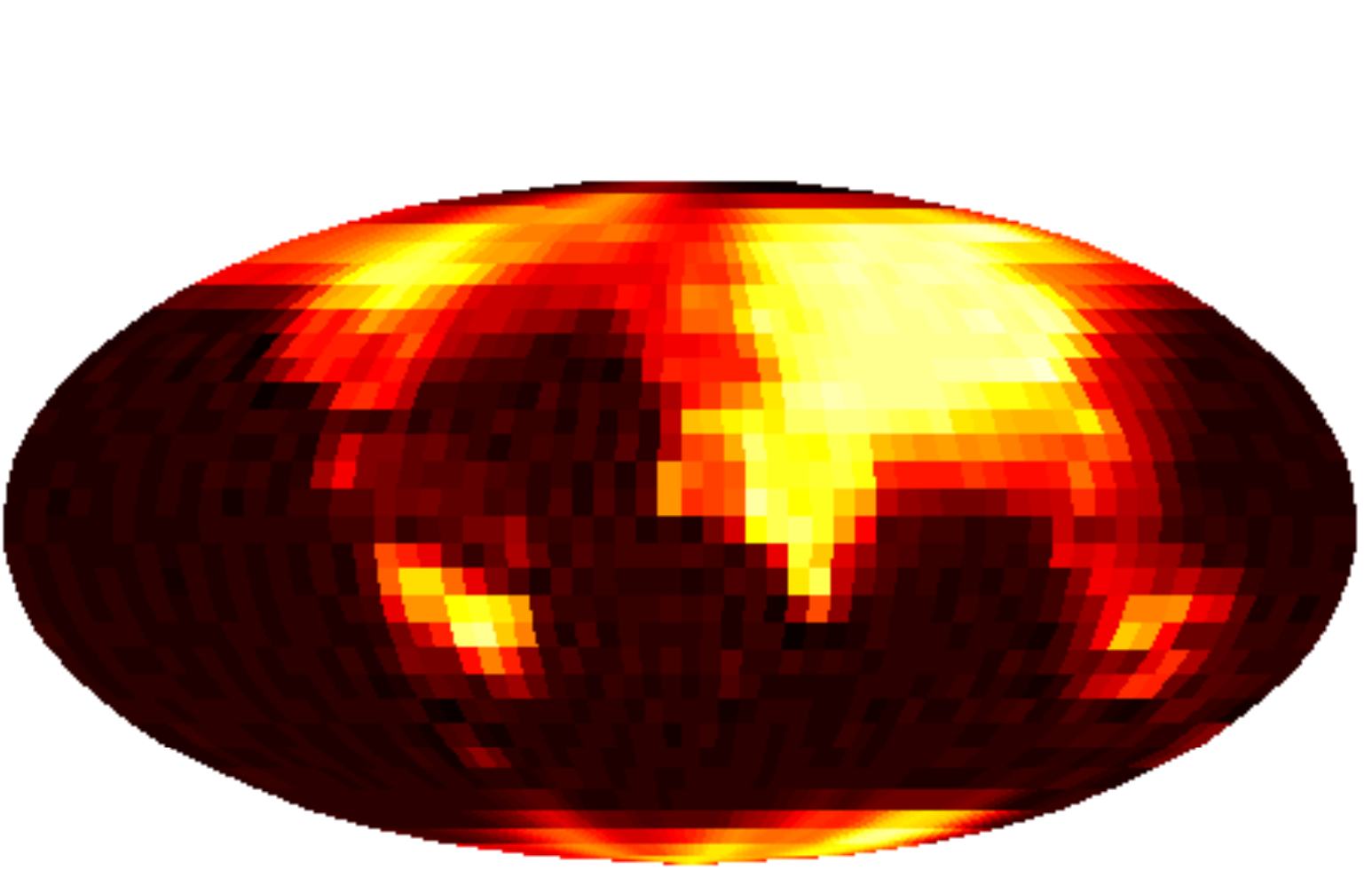}
  }
\subfigure[DH harmonic for $\nmeas/\elmax^2 = 1/2$]
  {
    \includegraphics[clip=,viewport=1 2 441 223,width=\sphereplotwidth]{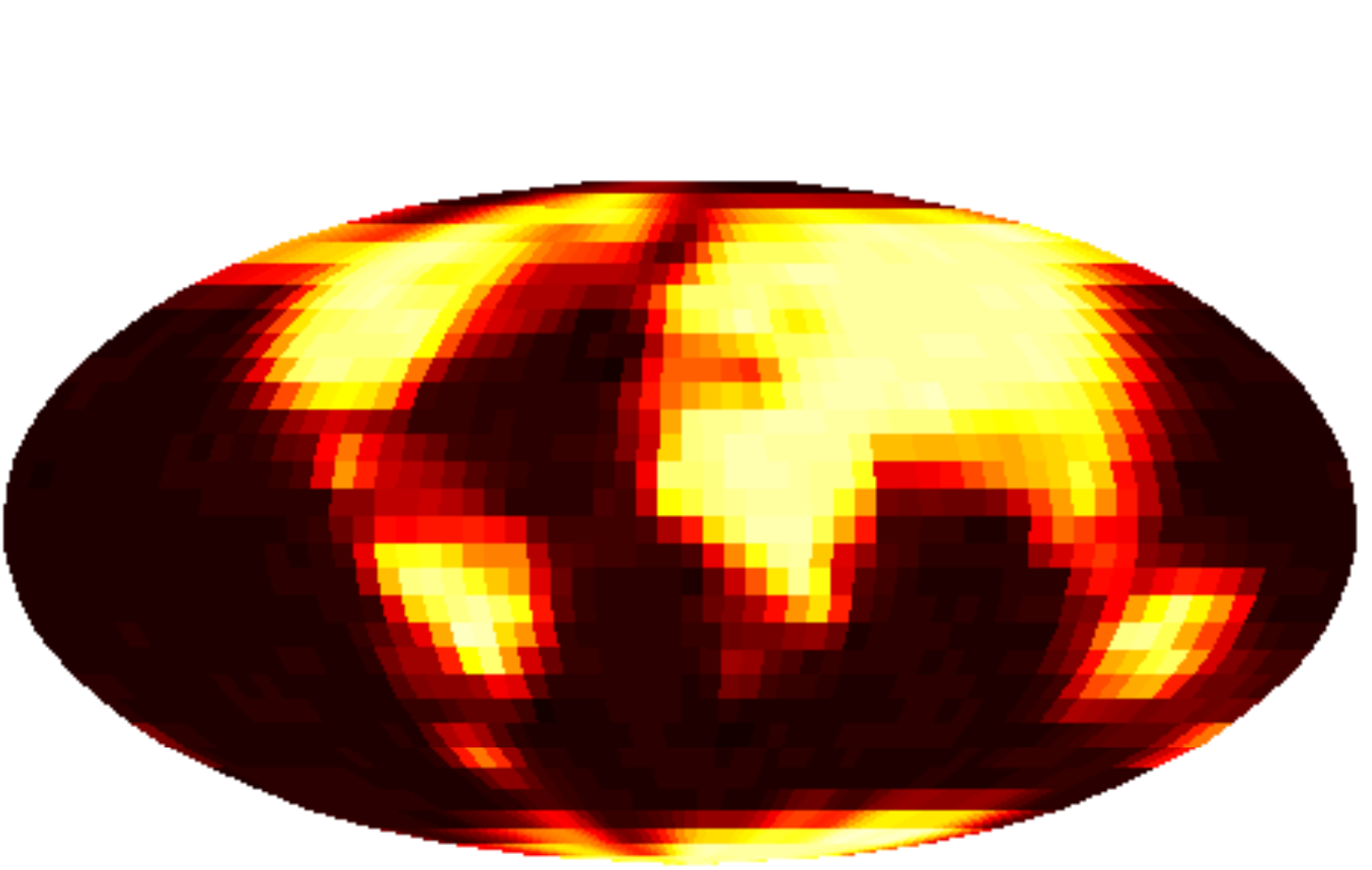}
  }
\subfigure[MW spatial for $\nmeas/\elmax^2 = 1/2$]
  {
    \includegraphics[clip=,viewport=1 2 441 223,width=\sphereplotwidth]{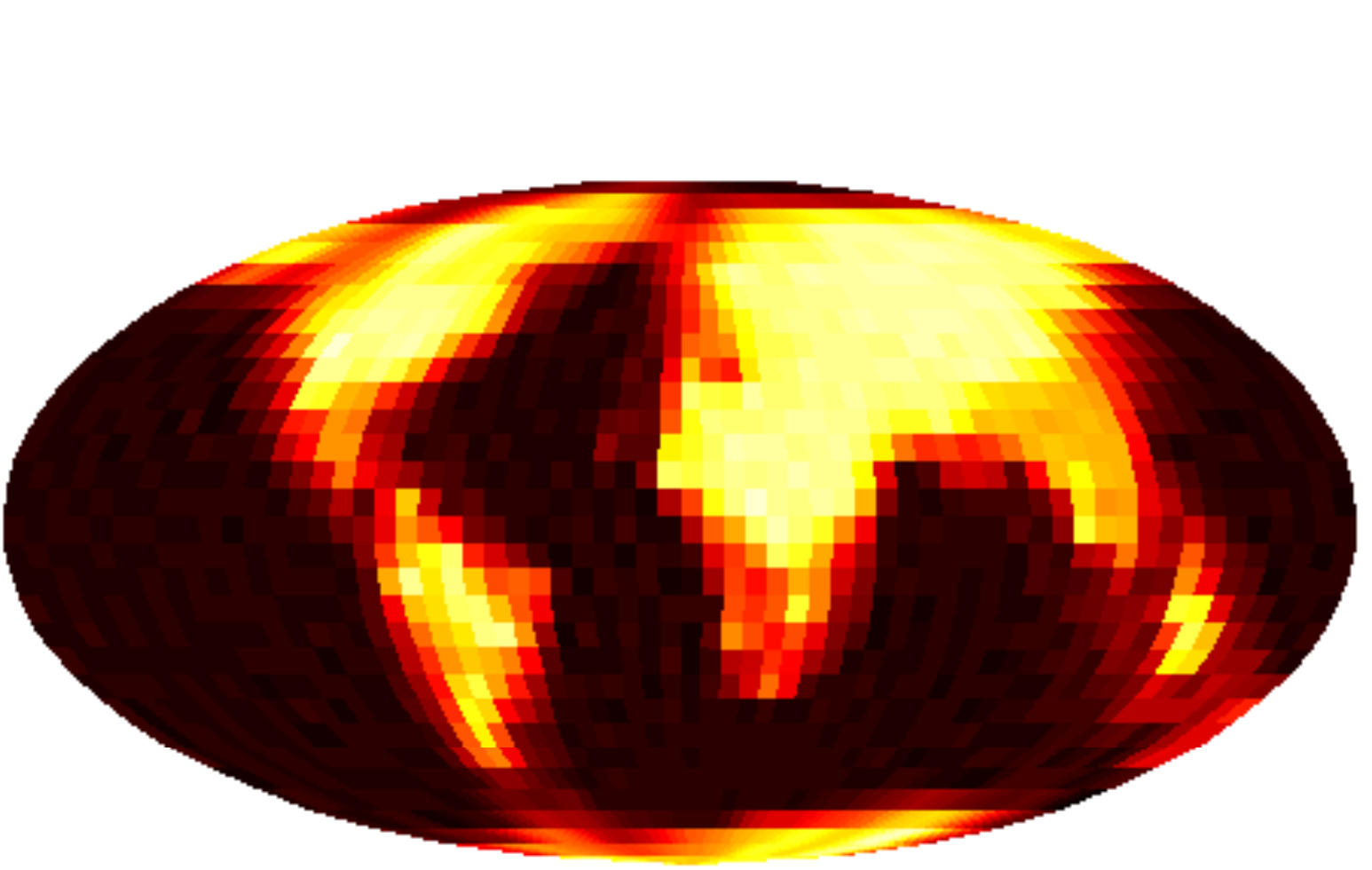}
  }
\subfigure[MW harmonic for $\nmeas/\elmax^2 = 1/2$]
  {
    \includegraphics[clip=,viewport=1 2 441 223,width=\sphereplotwidth]{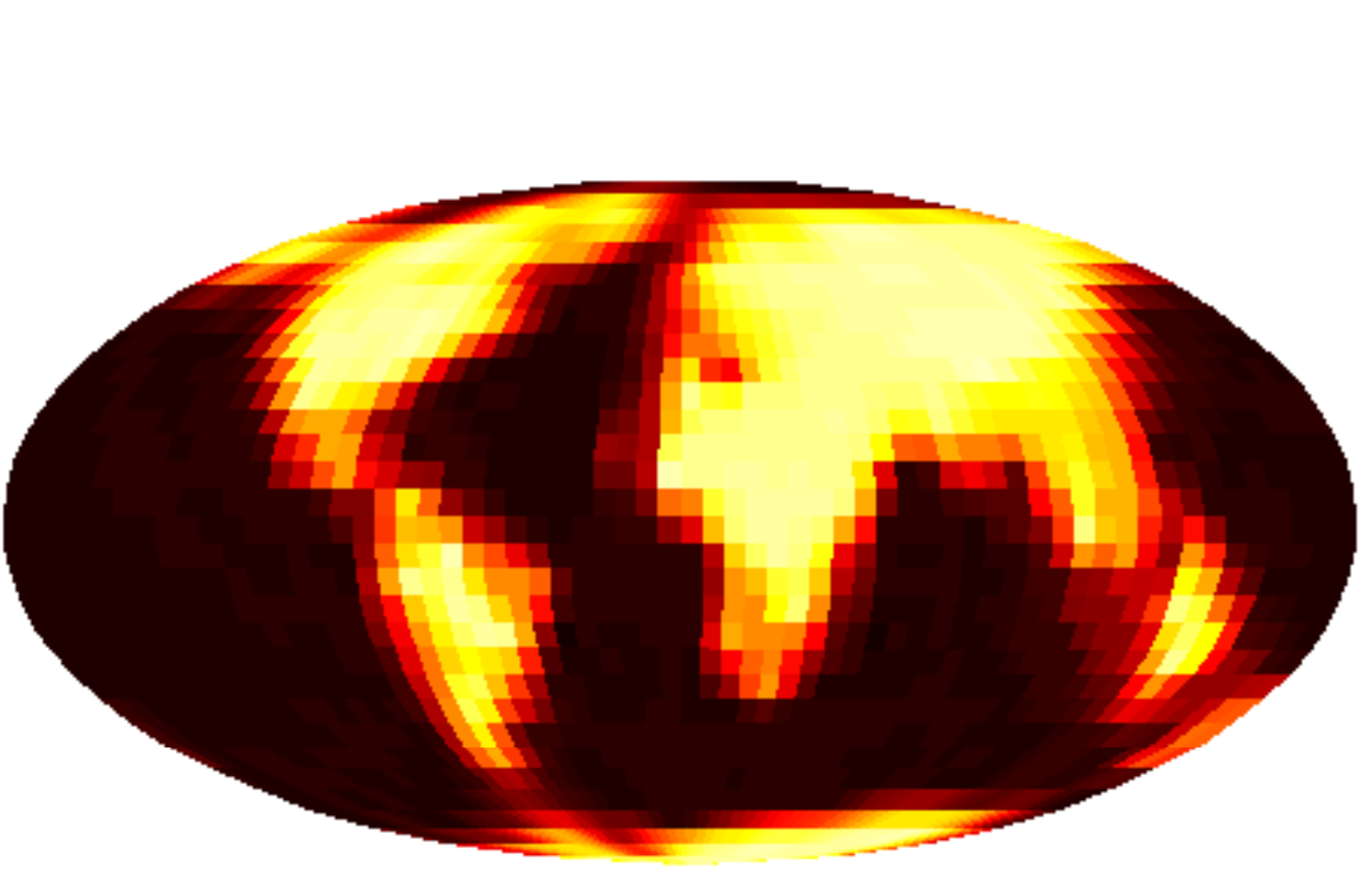}
  } 
}\\
\mbox{
\subfigure[DH spatial for $\nmeas/\elmax^2 = 1$]
  {
    \includegraphics[clip=,viewport=1 2 441 223,width=\sphereplotwidth]{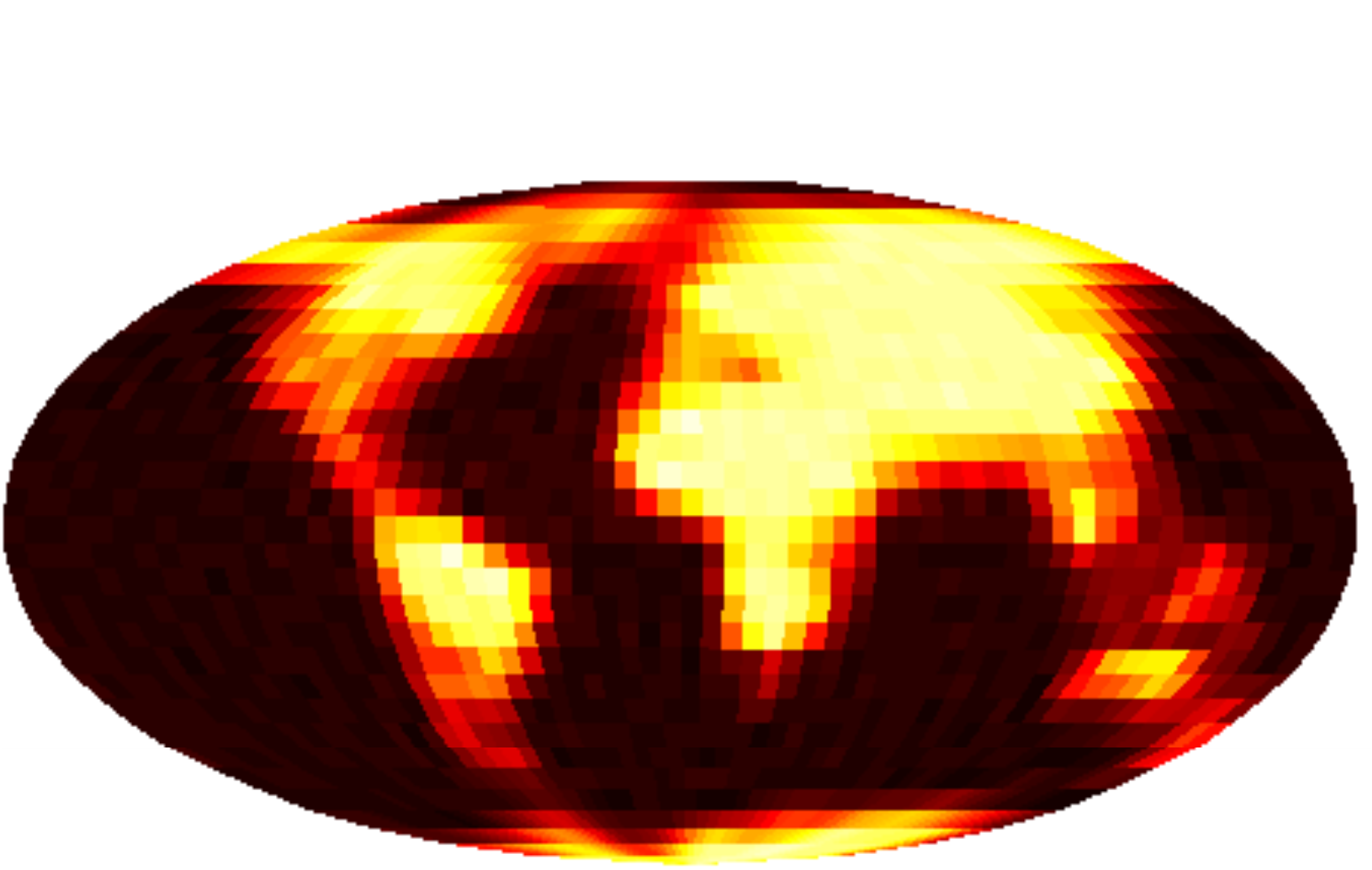}
   }
\subfigure[DH harmonic for $\nmeas/\elmax^2 = 1$]
  {
   \includegraphics[clip=,viewport=1 2 441 223,width=\sphereplotwidth]{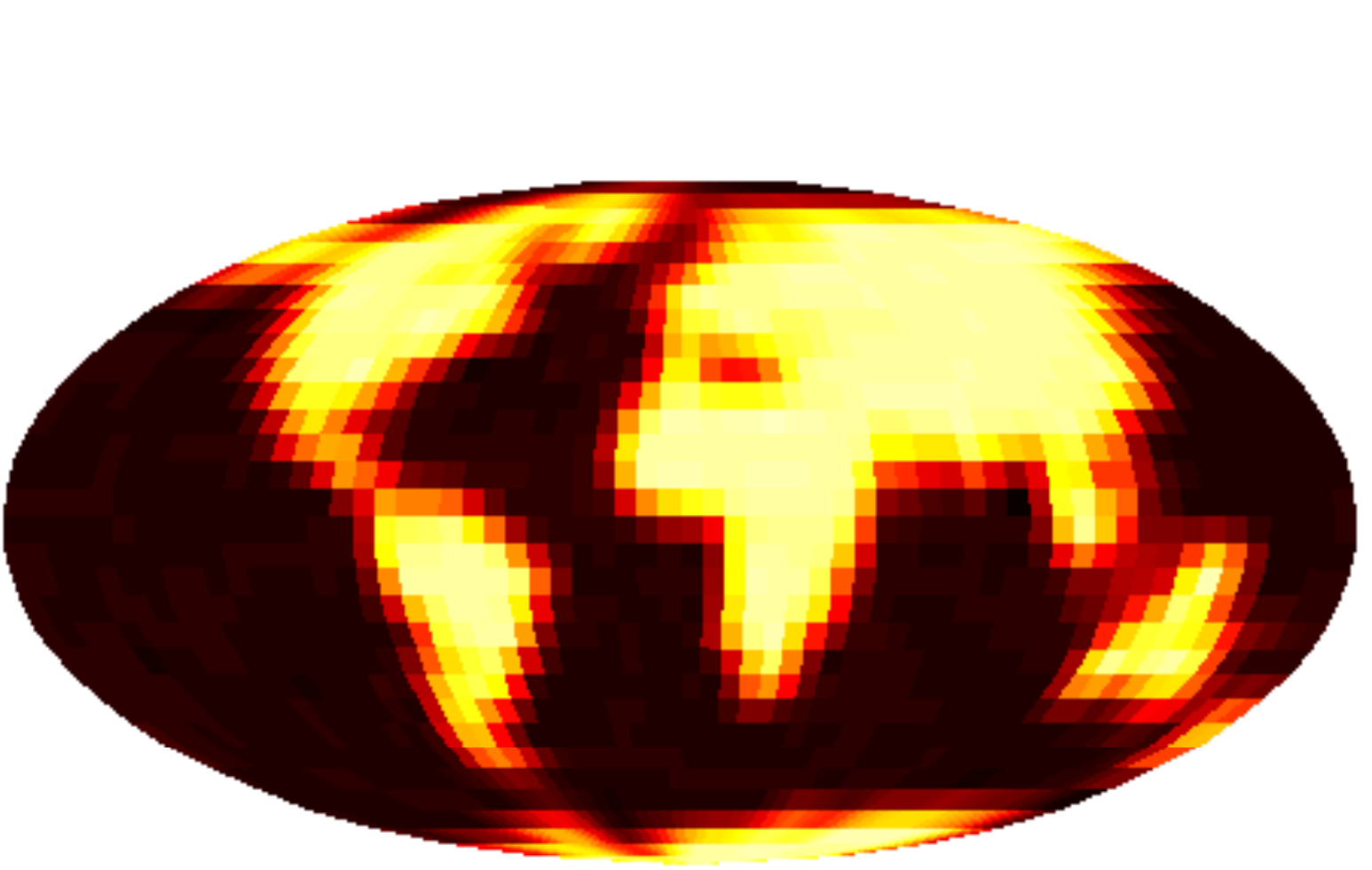}
  }
\subfigure[MW spatial for $\nmeas/\elmax^2 = 1$]
  {
    \includegraphics[clip=,viewport=1 2 441 223,width=\sphereplotwidth]{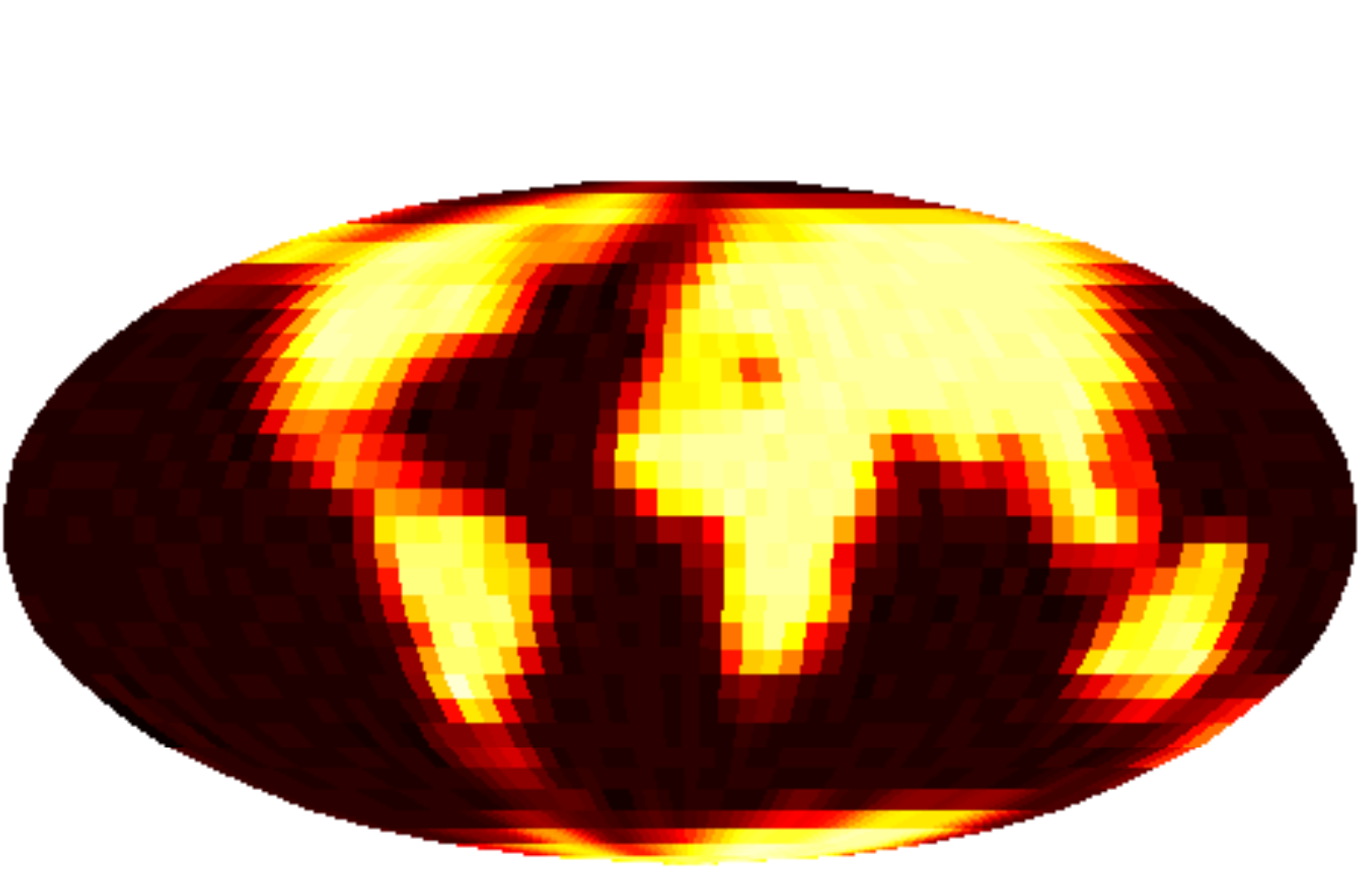}
  }
\subfigure[MW harmonic for $\nmeas/\elmax^2 = 1$]
  {
    \includegraphics[clip=,viewport=1 2 441 223,width=\sphereplotwidth]{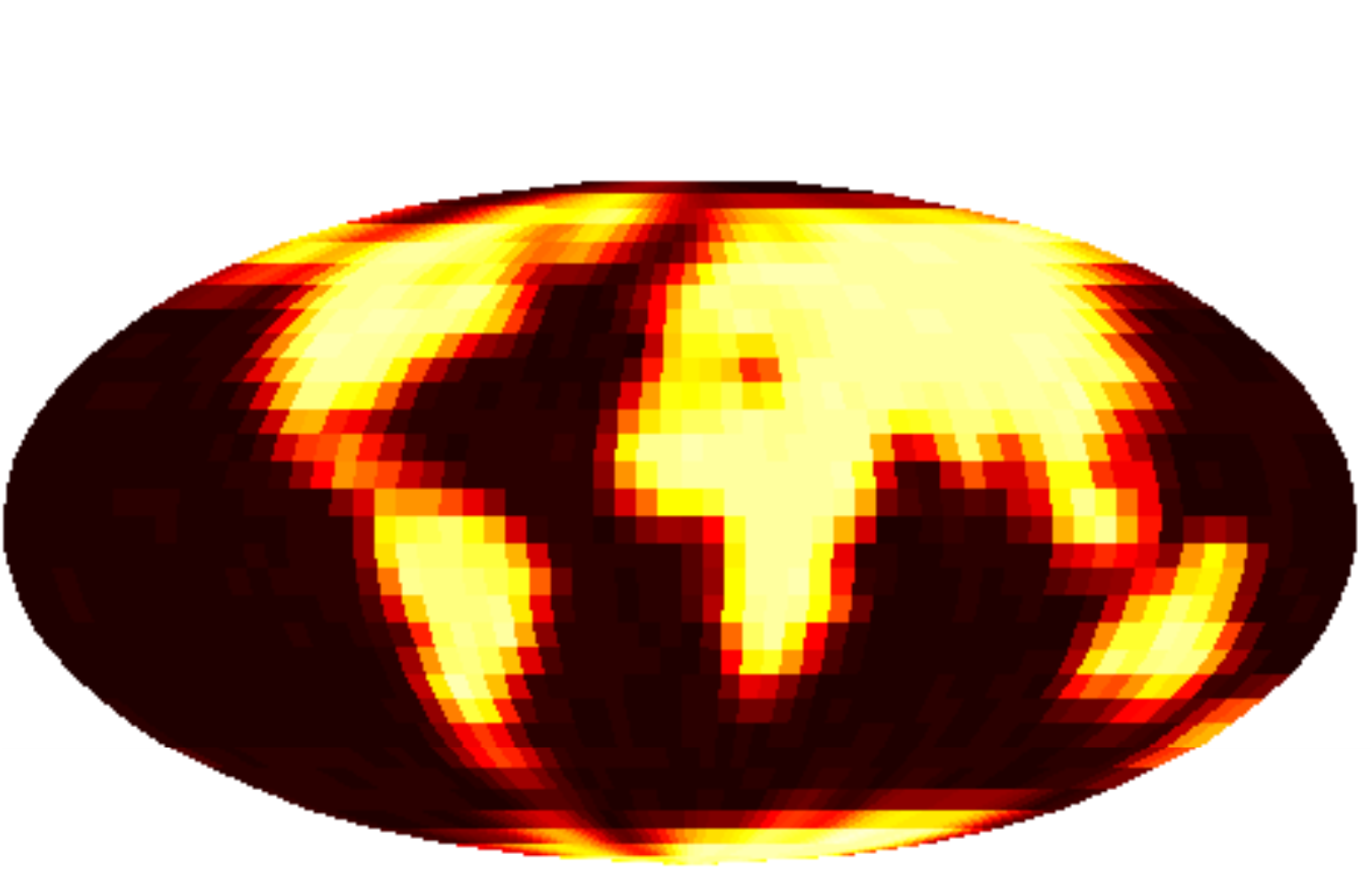}
  } 
}\\
\mbox{
\subfigure[DH spatial for $\nmeas/\elmax^2 = 3/2$]
  {
    \includegraphics[clip=,viewport=1 2 441 223,width=\sphereplotwidth]{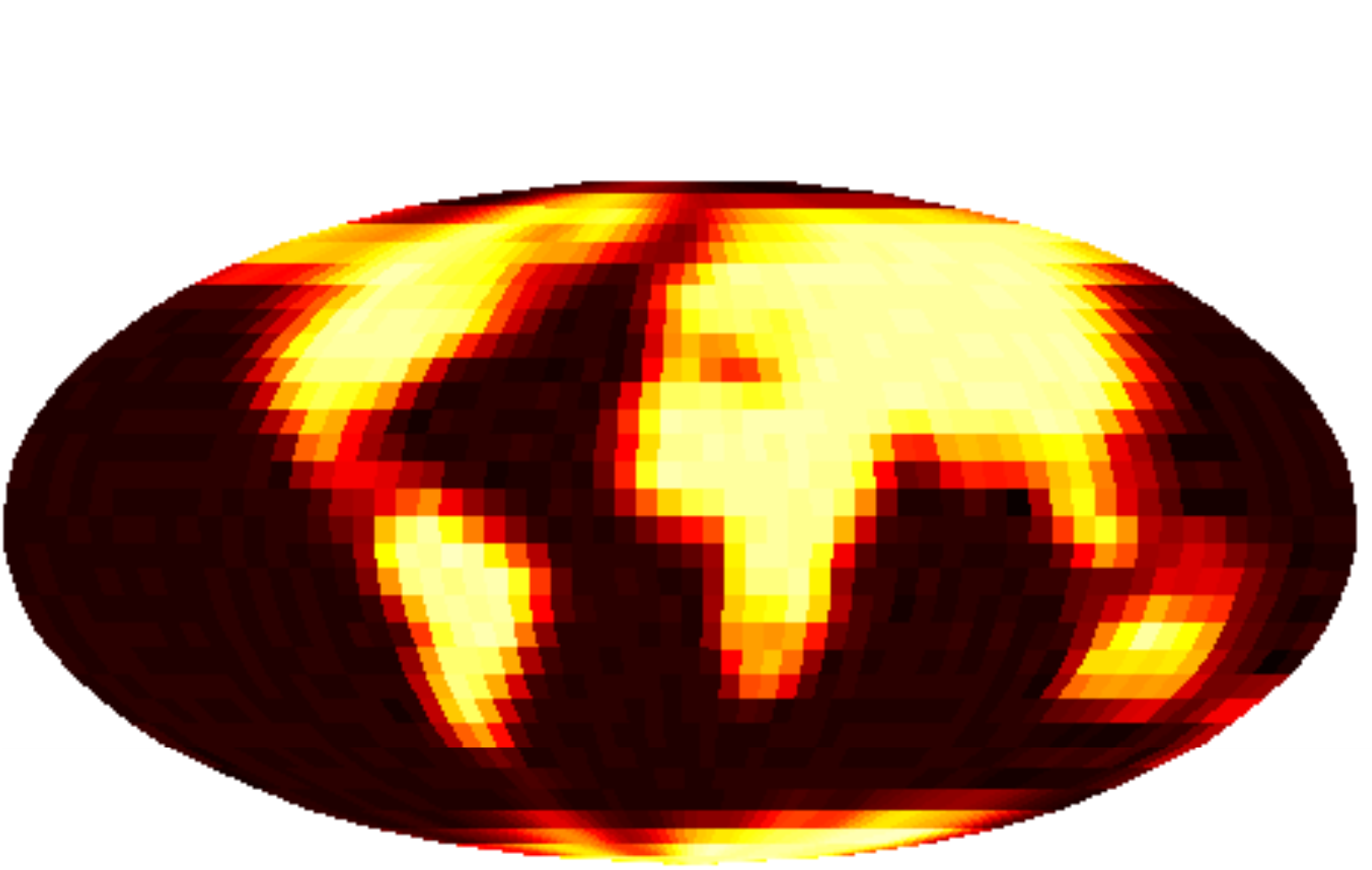}
  }
\subfigure[DH harmonic for $\nmeas/\elmax^2 = 3/2$]
  {
    \includegraphics[clip=,viewport=1 2 441 223,width=\sphereplotwidth]{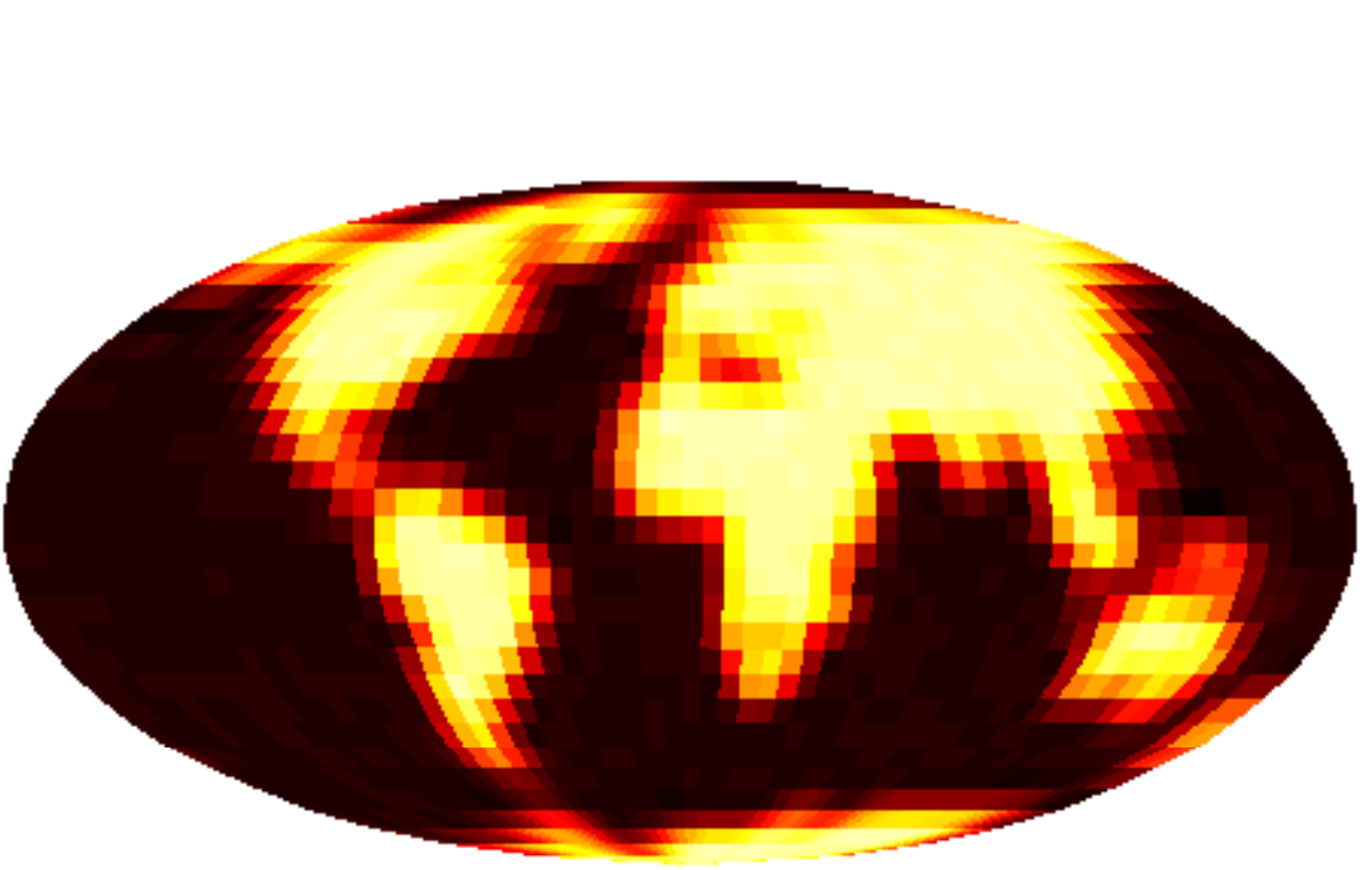}
  }
\subfigure[MW spatial for $\nmeas/\elmax^2 = 3/2$]
  {
    \includegraphics[clip=,viewport=1 2 441 223,width=\sphereplotwidth]{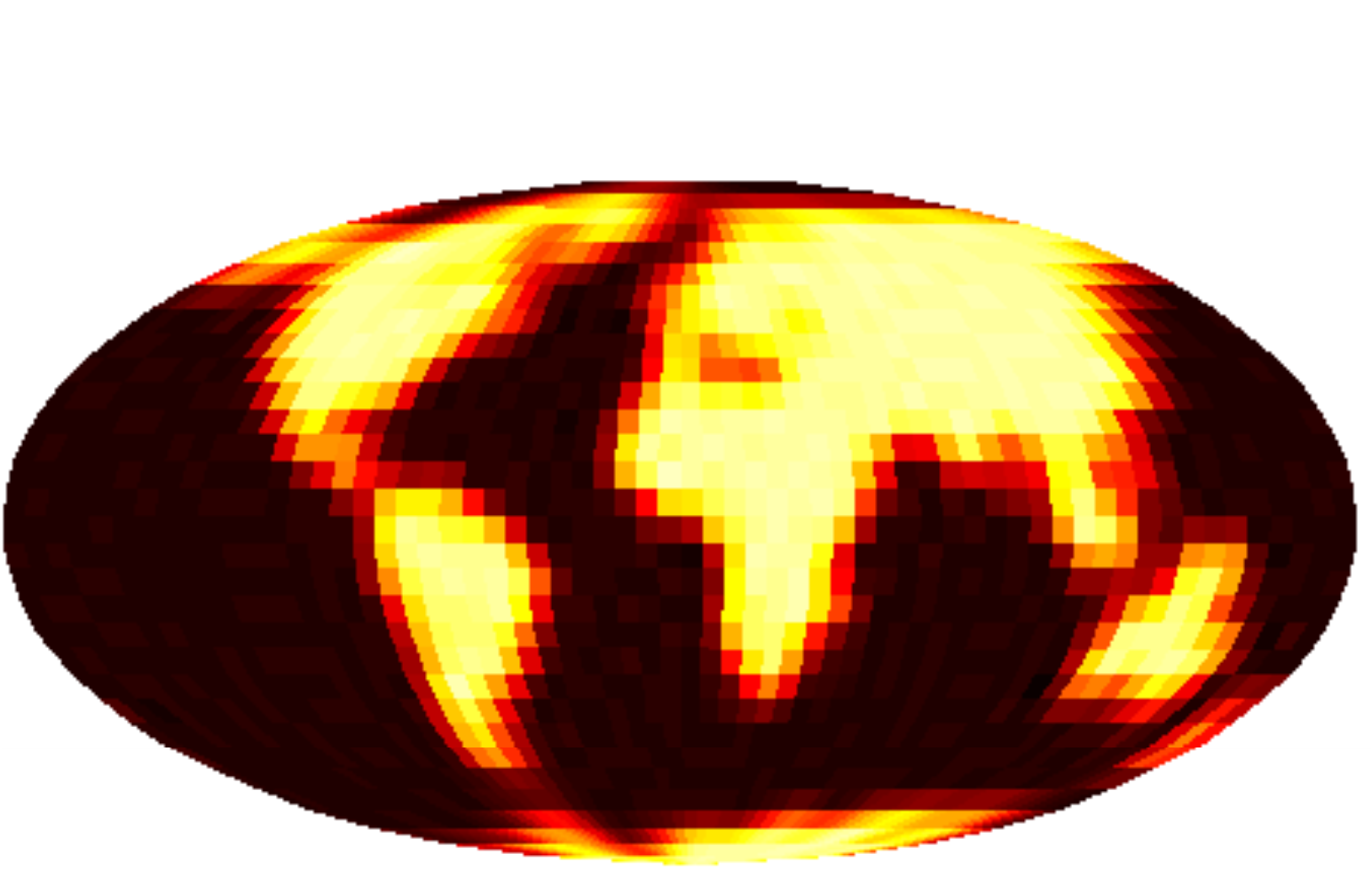}
  }
\subfigure[MW harmonic for $\nmeas/\elmax^2 = 3/2$]
  {
    \includegraphics[clip=,viewport=1 2 441 223,width=\sphereplotwidth]{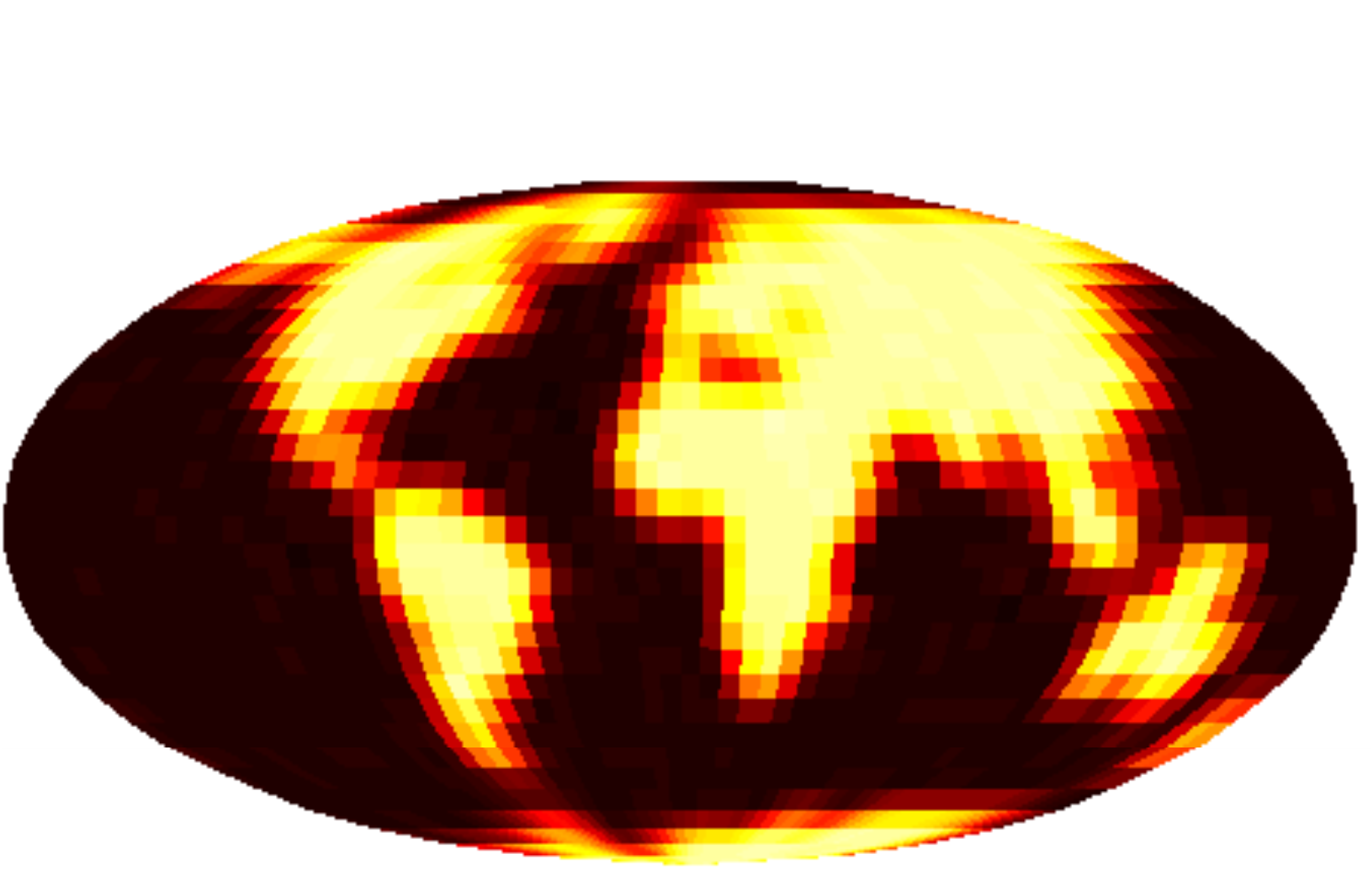}
  } 
}\\
\mbox{
\subfigure[DH spatial for $\nmeas/\elmax^2 \sim 2$]
  {
    \includegraphics[clip=,viewport=1 2 441 223,width=\sphereplotwidth]{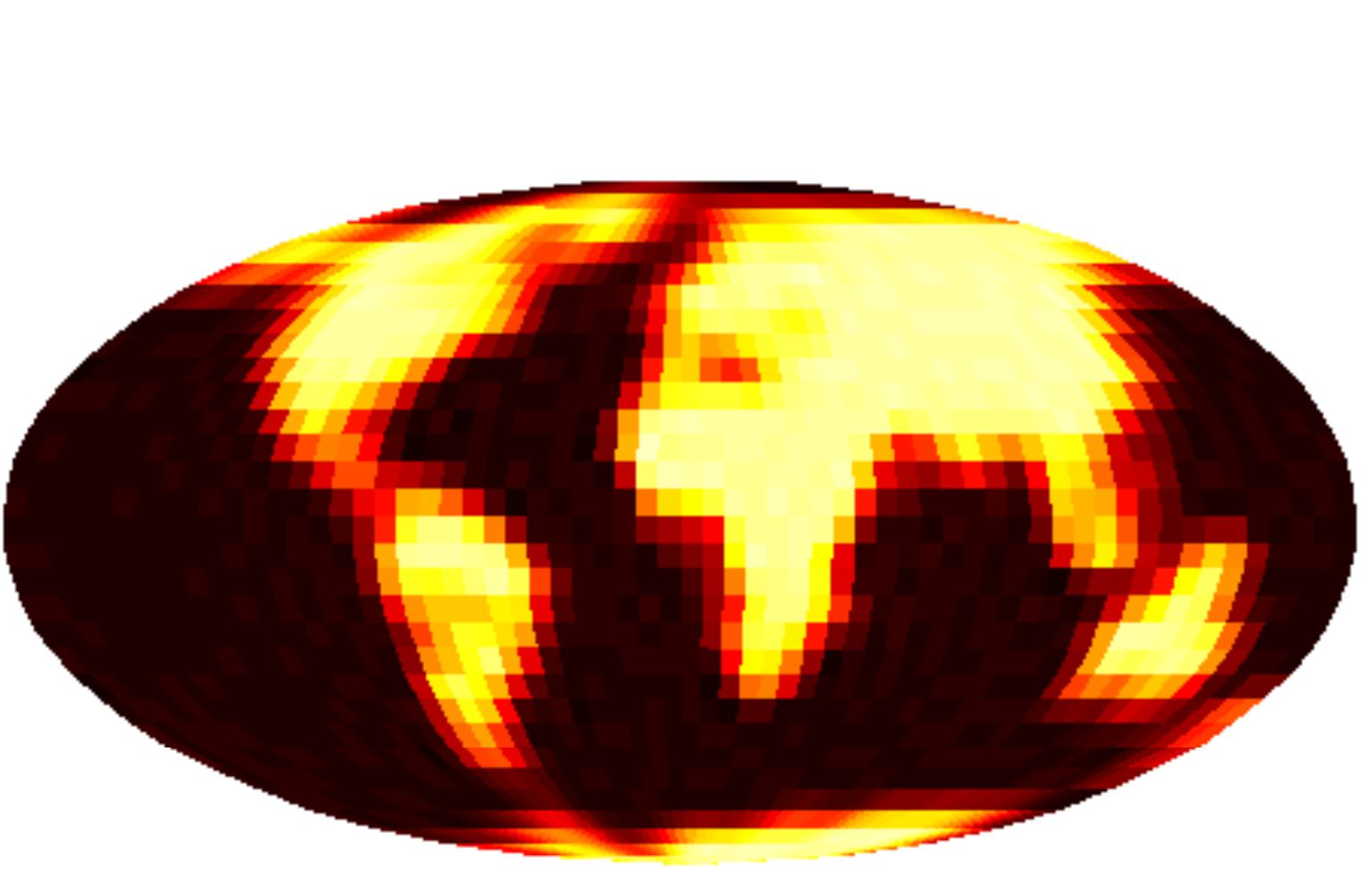}
  }
\subfigure[DH harmonic for $\nmeas/\elmax^2 \sim 2$]
  {
    \includegraphics[clip=,viewport=1 2 441 223,width=\sphereplotwidth]{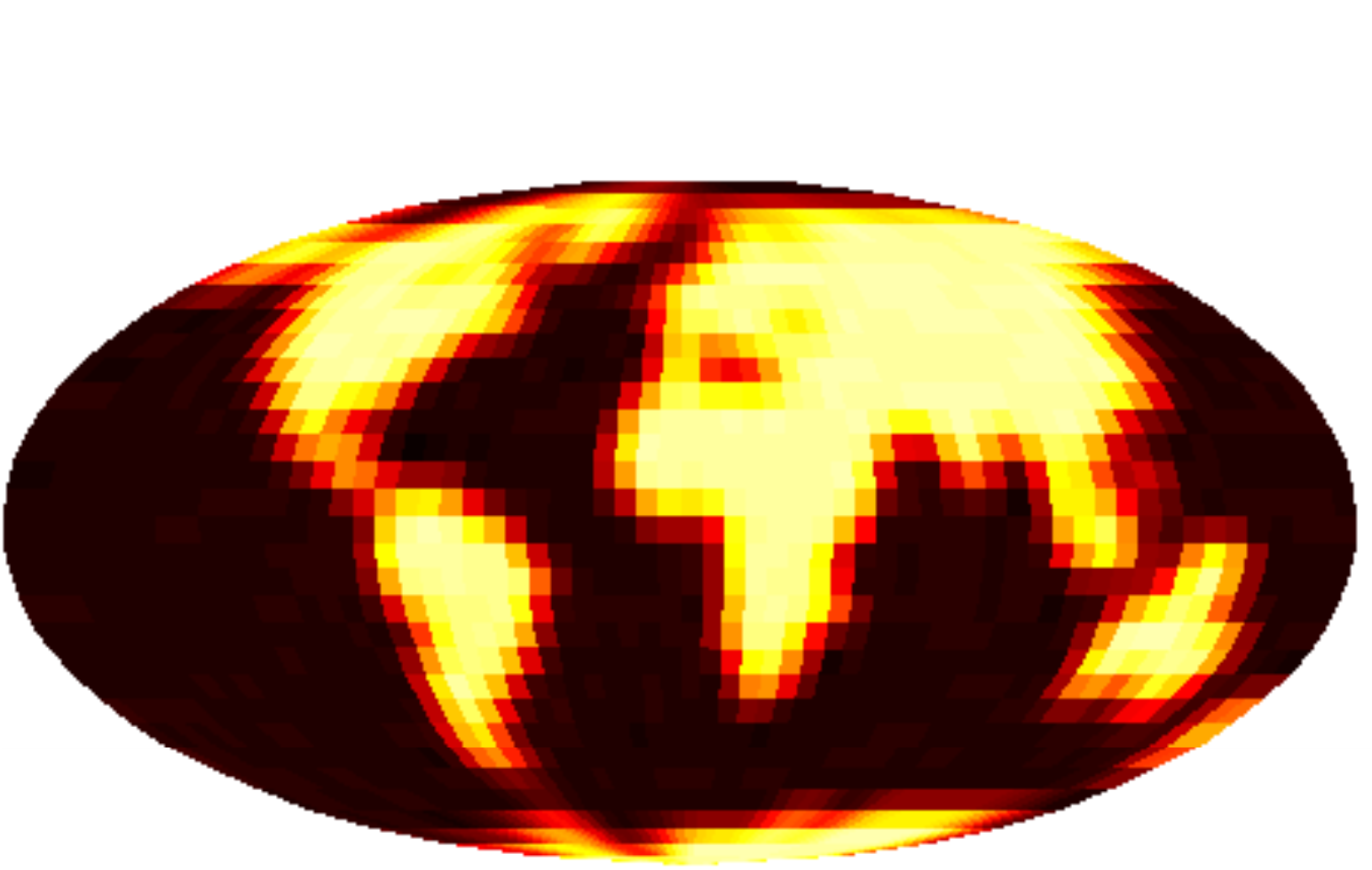}
  }
\subfigure[MW spatial for $\nmeas/\elmax^2 \sim 2$]
  {
    \includegraphics[clip=,viewport=1 2 441 223,width=\sphereplotwidth]{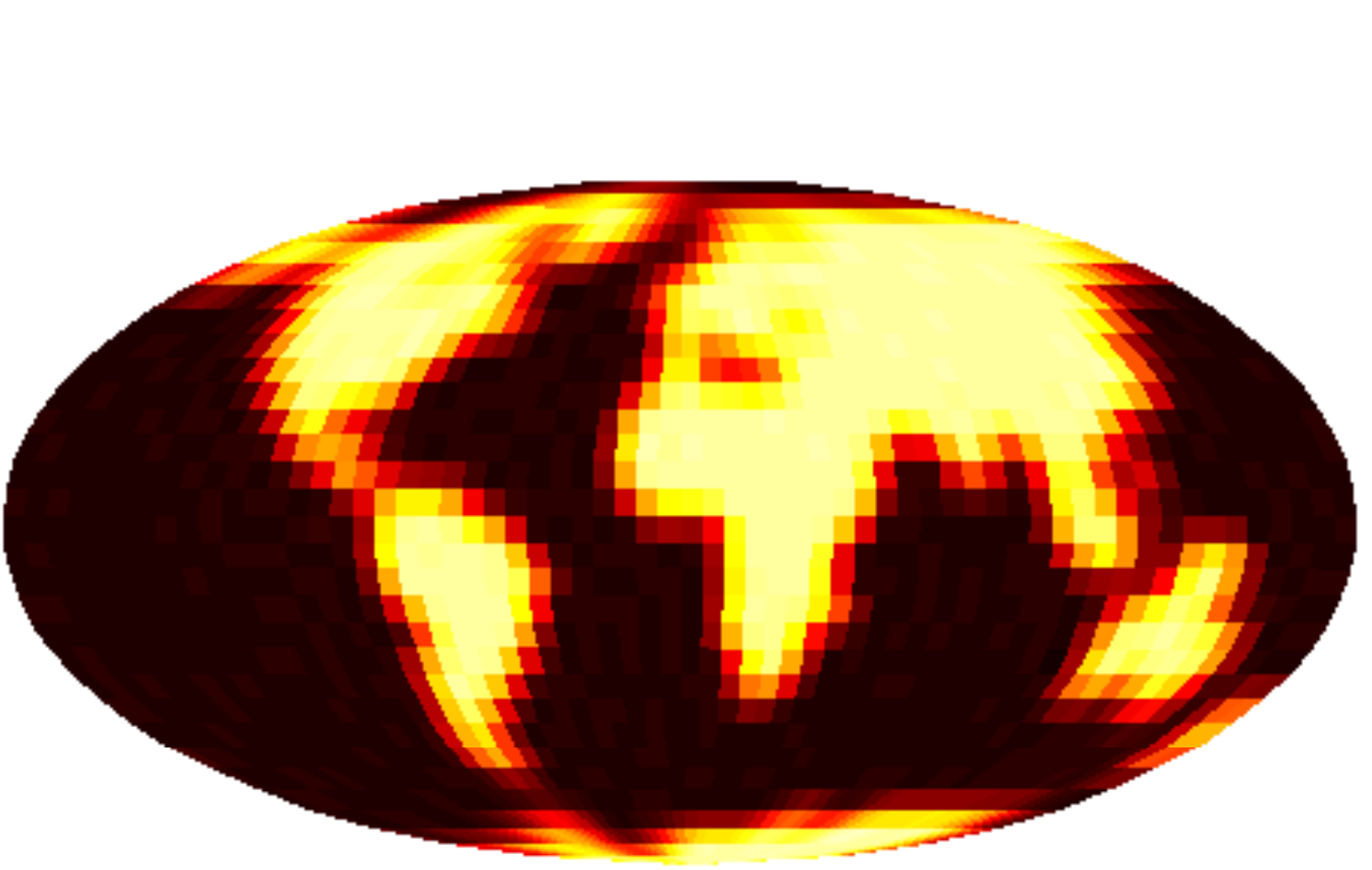}
  }
\subfigure[MW harmonic for $\nmeas/\elmax^2 \sim 2$]
  {
    \includegraphics[clip=,viewport=1 2 441 223,width=\sphereplotwidth]{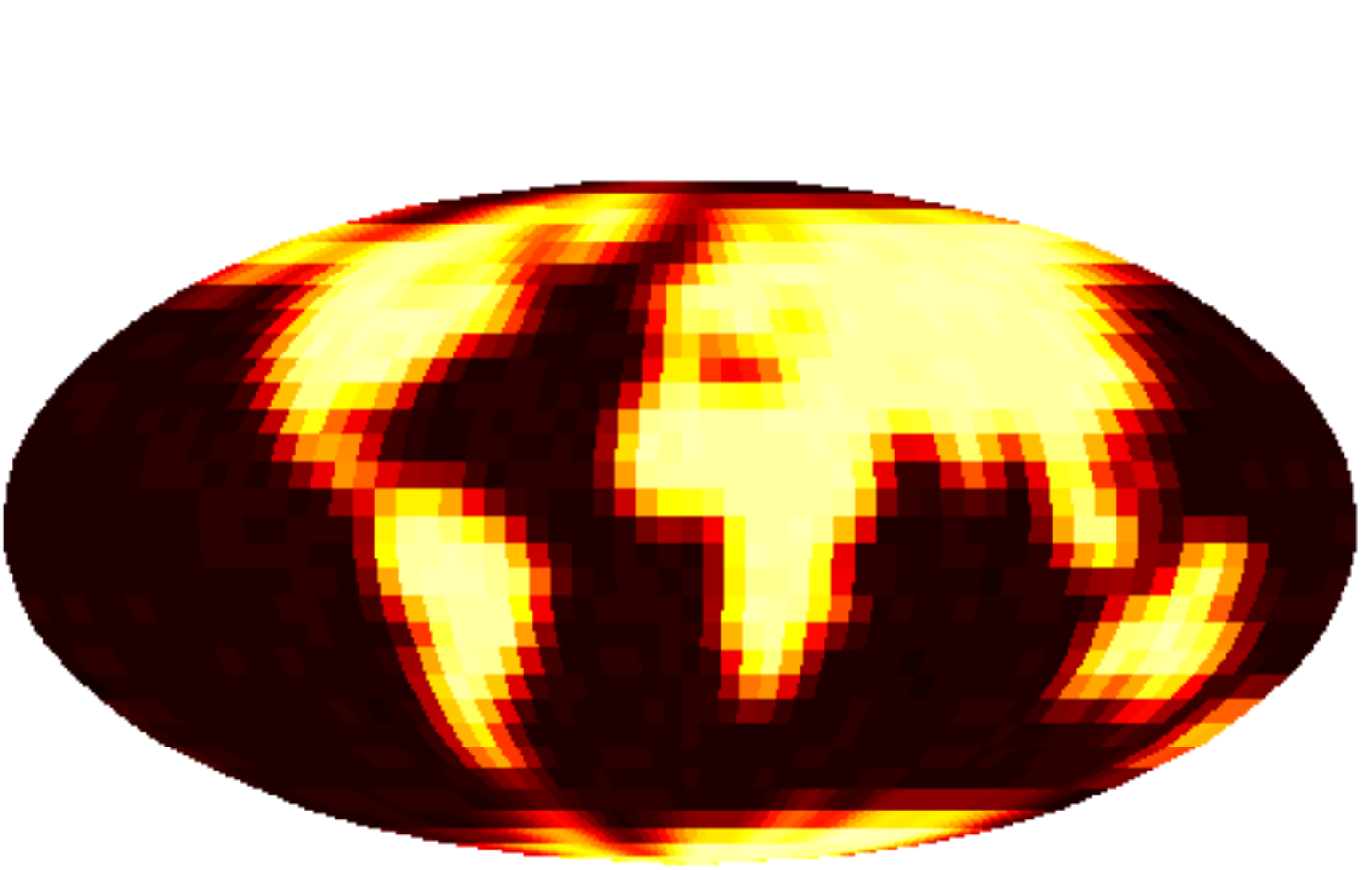}
  } 
}
\caption{Inpainted images on the sphere recovered by solving the TV
  inpainting problems for a range of measurement ratios
  $\nmeas/\elmax^2$. The ground truth image is shown in
    \fig{\ref{fig:truth}}. The first and second columns of panels show
  the inpainted images recovered using the DH sampling theorem, while
  the third and fourth columns show the inpainted images recovered
  using the MW sampling theorem.  The first and third columns of
  panels show inpainted images recovered by solving the inpainting
  problem in the spatial domain, while the second and fourth columns
  show images recovered by solving the inpainting problem in the
  harmonic domain.  The final row of panels corresponds to measurement
  ratio $\nmeas/\elmax^2 = \Nmw/\elmax^2 \sim 2$.  The quality
  enhancements due to the MW sampling theorem and by solving the
  inpainting problem in harmonic space are both clear.}
\label{fig:reconstructions}
\end{figure*}

%==============================================================================
\subsection{High-resolution illustration on a realistic image}
\label{sec:simulations:hires}

In this section we perform a single simulation to illustrate TV
inpainting at high resolution.  Furthermore, we also consider a
  more realistic test image.  Since we develop fast adjoint
algorithms for the MW sampling theorem only (due to its superiority),
we therefore use only the MW sampling theorem for the high-resolution
inpainting simulation performed here.

\begin{figure*}
\centering
\subfigure[Ground truth]{\includegraphics[clip=,viewport=1 0 445 223,width=70mm]{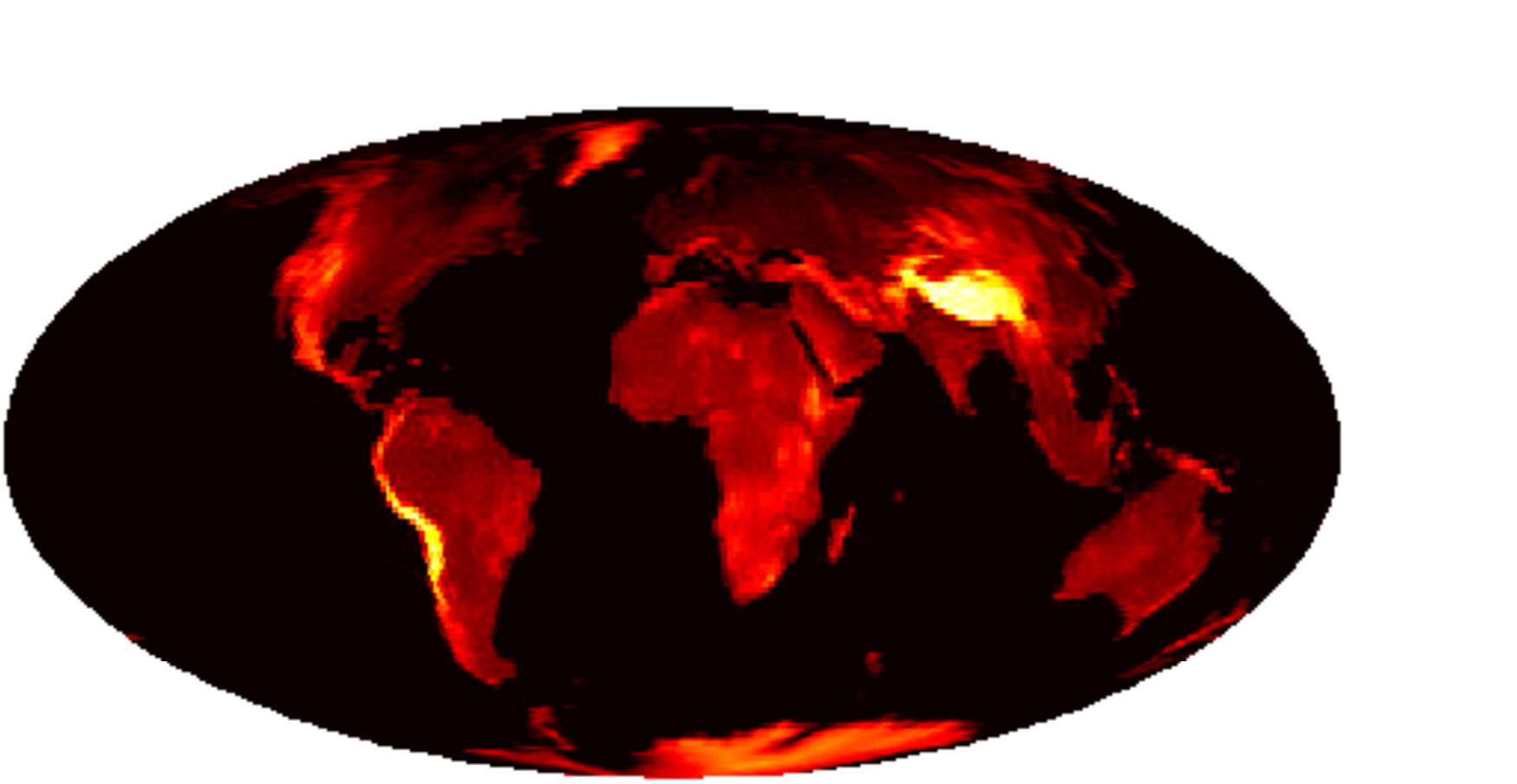}}\quad
\subfigure[Inpainted image for $\nmeas/\elmax^2 =1/4$ ($\snr_{\rm I}=20.0$dB)]{\includegraphics[clip=,viewport=1 0 445 223,width=70mm]{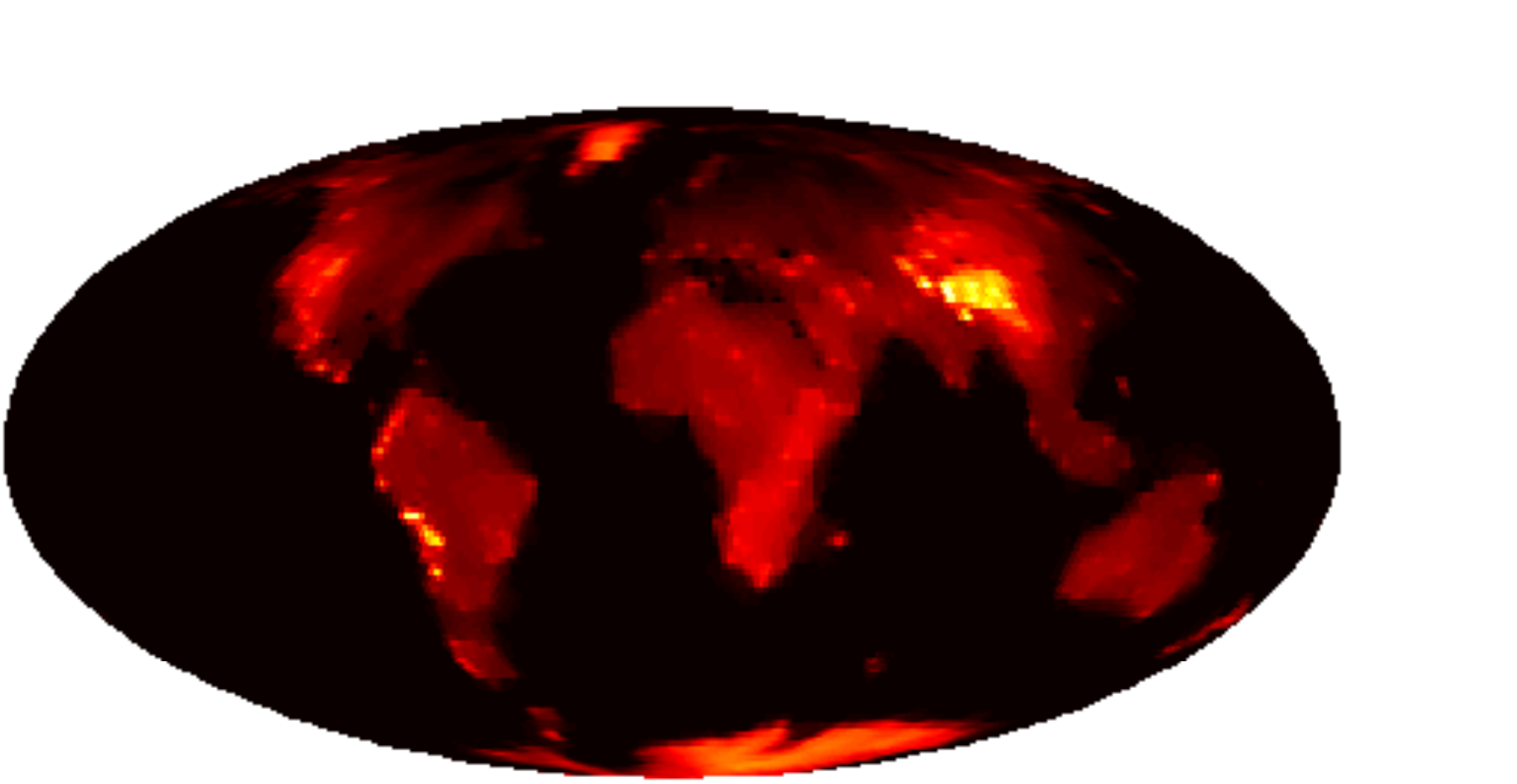}}
\subfigure[Inpainted image for $\nmeas/\elmax^2 =1/2$ ($\snr_{\rm I}=27.8$dB)]{\includegraphics[clip=,viewport=1 0 445 223,width=70mm]{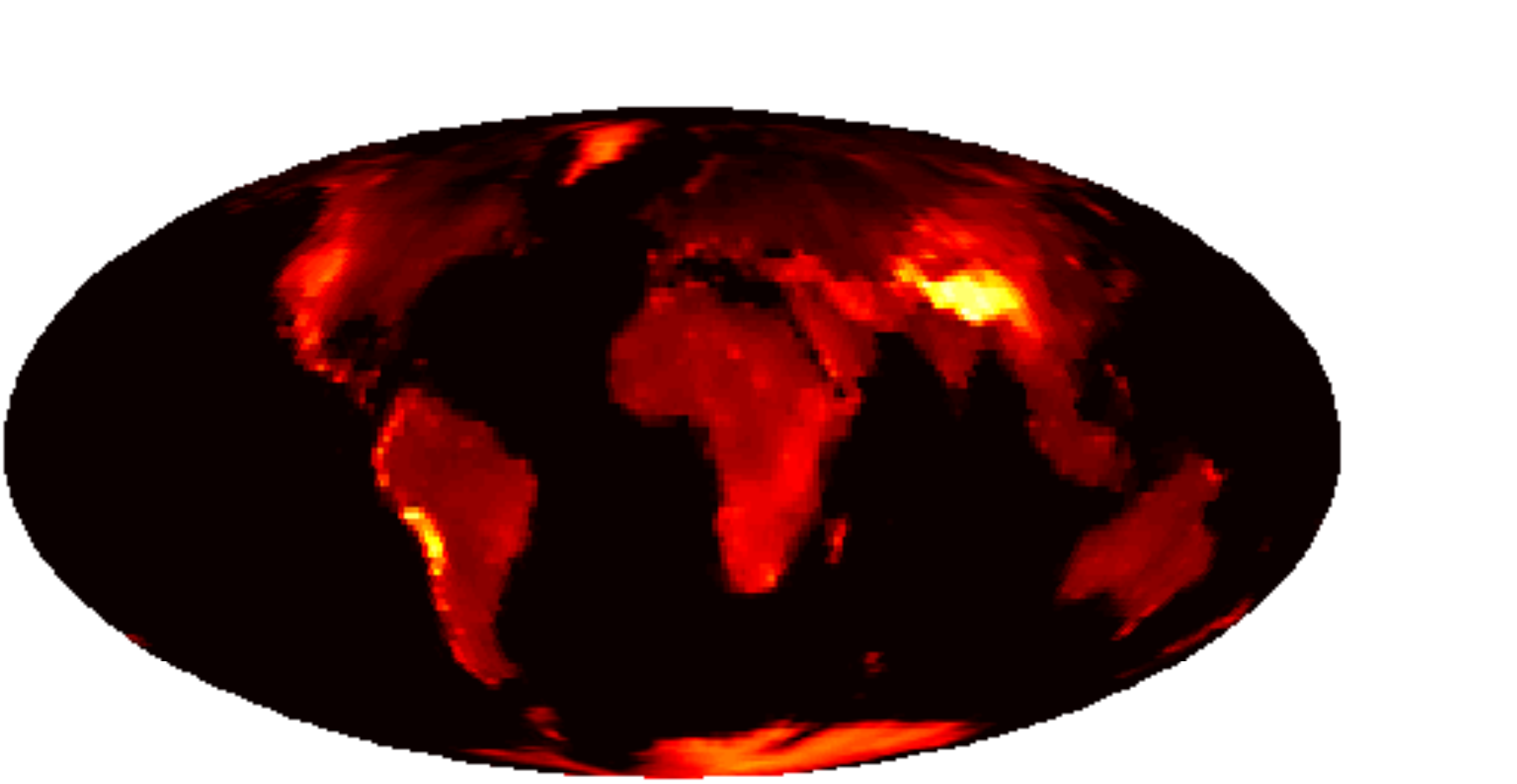}}\quad
\subfigure[Inpainted image for $\nmeas/\elmax^2 =1$ ($\snr_{\rm I}=37.0$dB)]{\includegraphics[clip=,viewport=1 0 445 223,width=70mm]{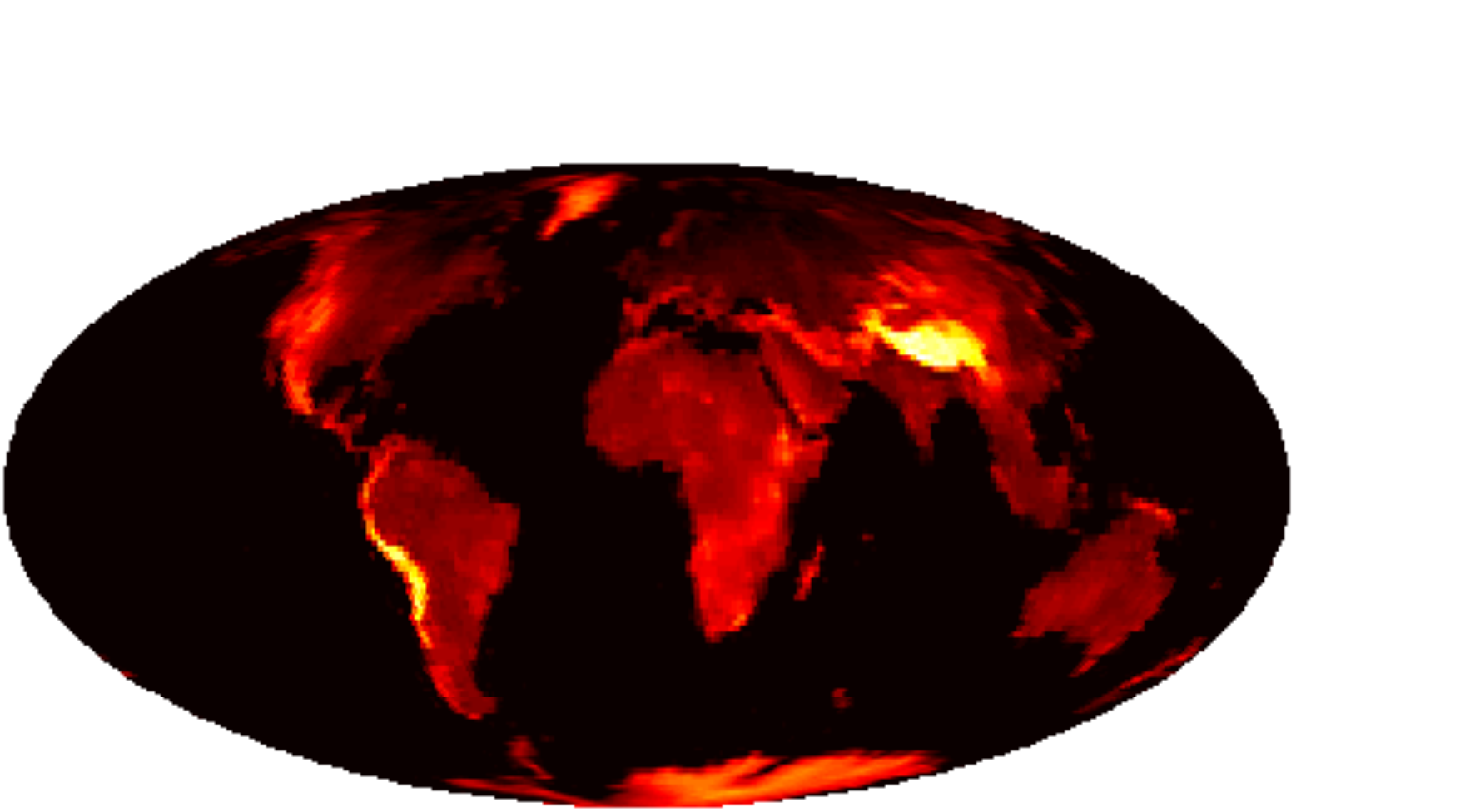}}
\subfigure[Inpainted image for $\nmeas/\elmax^2 = 3/2$ ($\snr_{\rm I}=38.5$dB)]{\includegraphics[clip=,viewport=1 0 445 223,width=70mm]{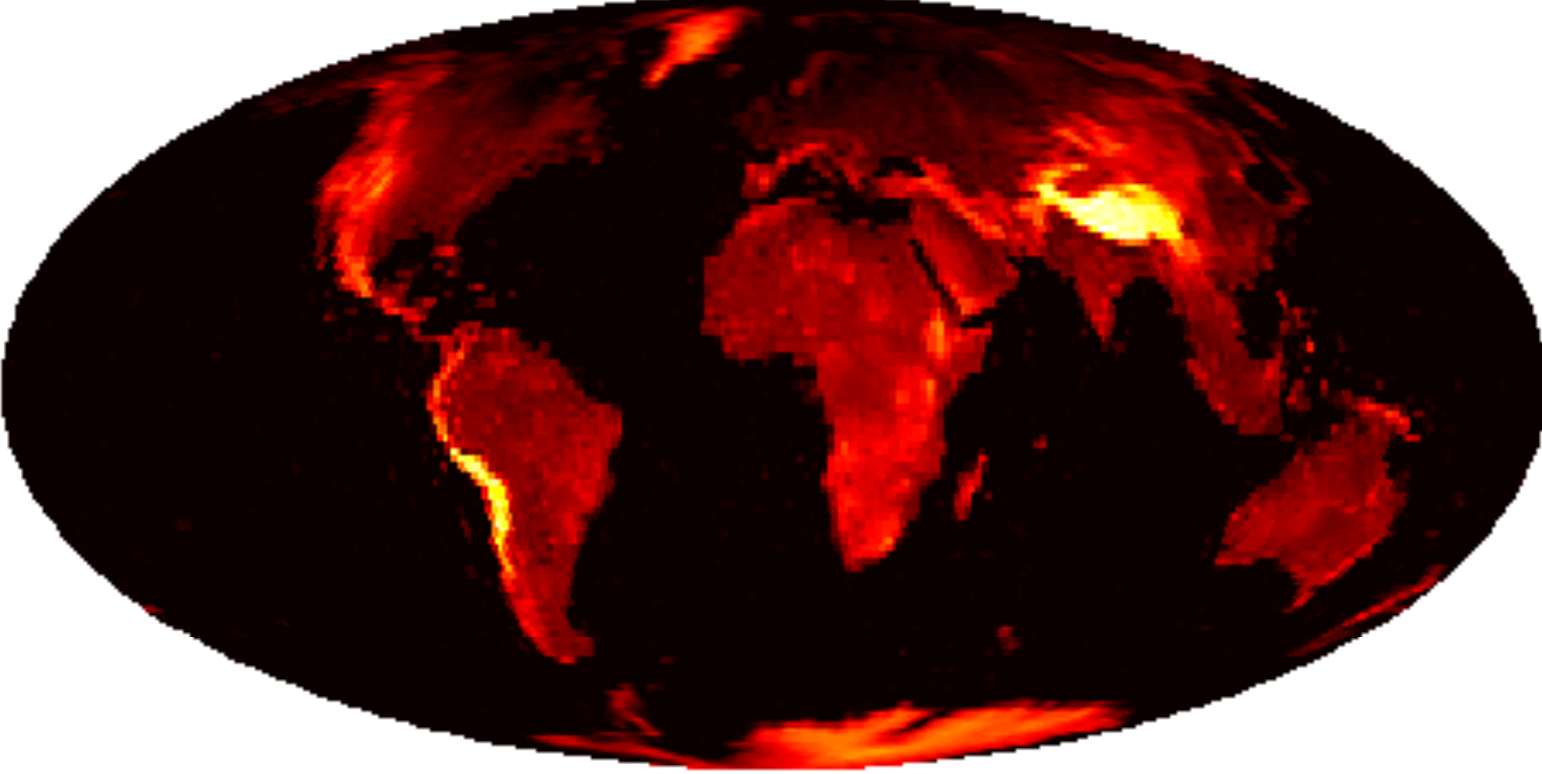}}\quad
\subfigure[Inpainted image for $\nmeas/\elmax^2 \sim 2$ ($\snr_{\rm I}=53.2$dB)]{\includegraphics[clip=,viewport=1 0 445 223,width=70mm]{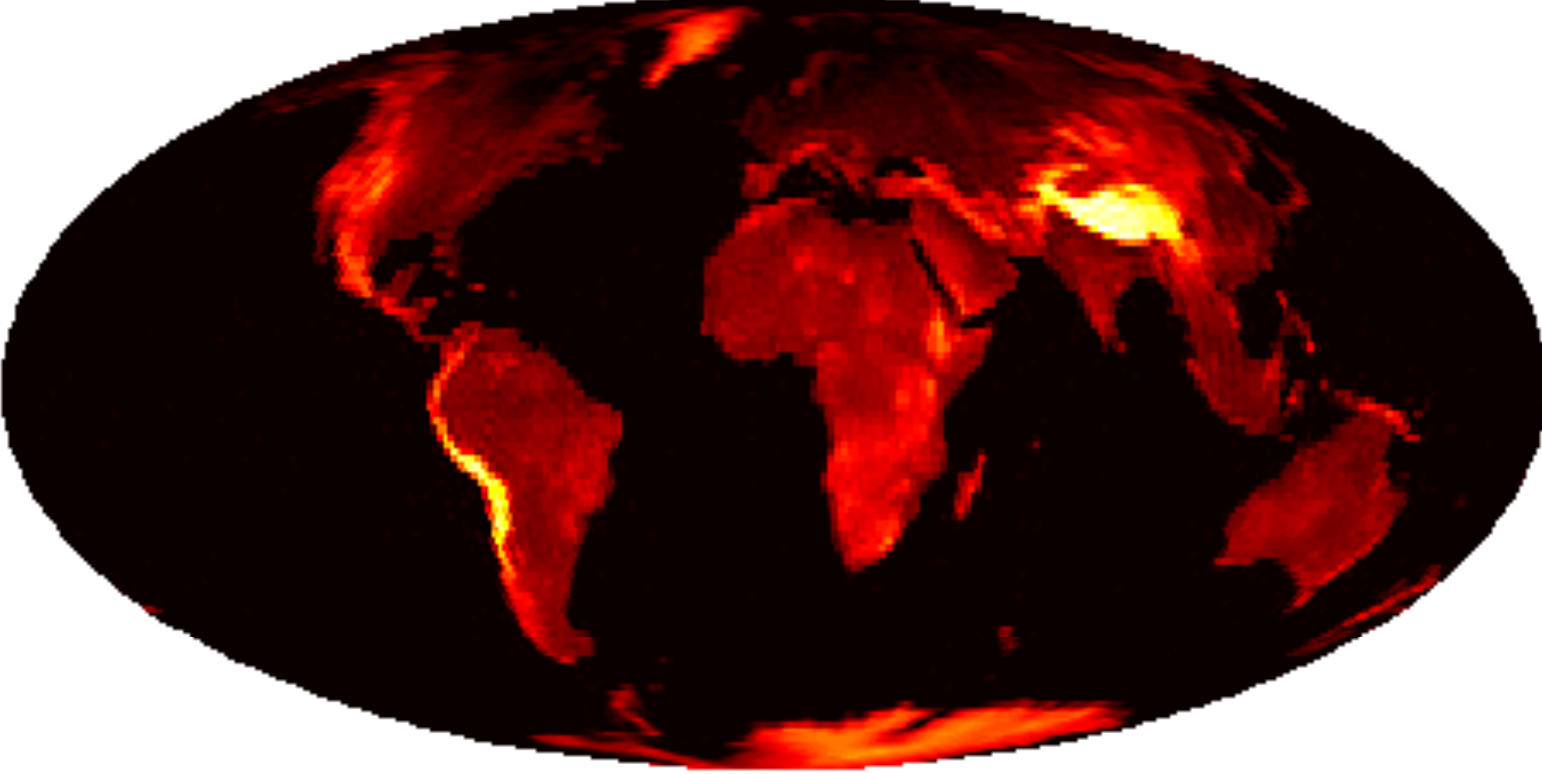}}
\caption{Inpainting illustration for a realistic image at
    high-resolution ($\elmax=128$). The inpainted images are recovered
    by solving the inpainting problem in harmonic space using the MW
    sampling theorem for a range of measurement ratios
    $\nmeas/\elmax^2$. The $\snr_{\rm I}$ of each recovered image is also
    displayed.}
\label{fig:sims_hires}
\end{figure*}

A high-resolution test image is constructed from the same Earth
topography data described in \sectn{\ref{sec:simulations:lores}}.
Since the original data are defined in harmonic space, we first simply
truncate the harmonic coefficients to yield an image band-limited at
$\elmax=128$.  In practice, acquired images may not necessarily be
band-limited.  We thus process the data to construct a test image that
is not band-limited.  We construct such a test image defined by height
above sea-level (\ie\ we threshold all oceans and trenches, while the
continents and mountains remain unaltered).  The abrupt transition
between the oceans and the continents results in an image that is
indeed not band-limited (as verified numerically).  Furthermore, the
continents and mountainous regions result in a test image that is not
highly sparse in its gradient.  The resulting realistic test image is
shown in \fig{\ref{fig:sims_hires}~(a)}.

The same measurement procedure as outlined previously is applied to
take noisy, incomplete measurements of the data for a range of
measurement ratios \mbox{$\nmeas/\elmax^2$}.  The inpainted images are
recovered by solving the inpainting problem in harmonic space through
\eqn{\ref{eqn:recon_harmonic}} using the MW sampling theorem.  To
solve the inpainting problem for these high-resolution simulations we
use the estimator of the inverse transform norm $\| \Lambda \|_2$
described in \sectn{\ref{sec:algorithms:bound}} and the fast adjoint
harmonic transform algorithms defined in
\sectn{\ref{sec:algorithms:adjoint}}.  Using these fast algorithms,
combined with recent optimisations of the {\tt SSHT} package, it takes
approximately 10 minutes to solve the inpainting problem in harmonic
space at $\elmax=128$ on a standard laptop (with a 1.8 GHz Intel Core
i7 processor and 4 GB of RAM).

The inpainted images are shown in \fig{\ref{fig:sims_hires}}. Since
the original realistic test image is not band-limited, the previous
\snr\ measure (which is defined in harmonic space to avoid a
dependence on the number of samples of each sampling theorem) is not a
meaningful error metric (computation of the harmonic transform would
indeed be affected by uncontrolled aliasing).  Instead, we use the
analogous \snr\ measure defined in image space on the sphere, given by
\mbox{$\snr_{\rm I}=10 \log ( \vect{{x}}^\dagger Q \vect{x} /
  ((\vect{{x}}^\star - \vect{{x}})^\dagger Q (\vect{{x}}^\star -
  \vect{{x}})))$}, where we recall $Q$ is the matrix with quadrature
weights on its diagonal.  Just as for the definition of the TV norm,
the inclusion of the weights for signals that are not band-limited
provides only an intuitive approximation to continuous
integration. The $\snr_{\rm I}$ values for each inpainted image are
displayed in \fig{\ref{fig:sims_hires}}, from which it is apparent
that $\snr_{\rm I}$ increases with increasing measurement ratio.
Moreover, it is clearly apparent by eye that our TV inpainting
framework is effective when applied to realistic images that are not
highly sparse in their gradient and that are not band-limited.

%==============================================================================
\section{Conclusions}
\label{sec:conclusions}
%==============================================================================

The MW sampling theorem, developed only recently, achieves a more
efficient sampling of the sphere than the standard DH sampling
theorem: without any loss to the information content of the sampled
signal, the MW sampling theorem reduces the number of samples required
to represent a band-limited signal by a factor of two for an
equiangular sampling.  For signals sparse in a spatially localised
measure, such as in a wavelet basis, overcomplete dictionary, or in
the magnitude of their gradient, for example, a more efficient
sampling enhances the fidelity of sparse image reconstruction through
both dimensionality and sparsity.  When a signal is recovered directly
in the spatial domain, the MW sampling theorem provides enhancements
in both dimensionality and sparsity when compared to the DH sampling
theorem.  By recovering the signal directly in harmonic space it is
possible to optimise its dimensionality, in which case the MW sampling
theorem still provides an enhancement in sparsity but not in
dimensionality.  

We verified these statements through a simple inpainting problem on
the sphere, where we considered images sparse in their gradient.  We
built a framework and fast methods for total variation (TV) inpainting
on the sphere.  Using this framework we performed numerical
experiments which confirmed our predictions: in all cases, the more
efficient sampling provided by the MW sampling theorem improved the
fidelity of sparse image reconstruction on the sphere.

\bibliographystyle{IEEEtran}
\bibliography{bib}

\begin{biographynophoto}{Jason McEwen}
  received a B.E.\ (Hons) degree in Electrical and Computer
  Engineering from the University of Canterbury, New Zealand, in 2002
  and a Ph.D.\ degree in Astrophysics from the University of Cambridge
  in 2006.

  He held a Research Fellowship at Clare College, Cambridge, from 2007
  to 2008, worked as a Quantitative Analyst from 2008 to 2010, and
  held a position as a Postdoctoral Researcher at Ecole Polytechnique
  F{\'e}d{\'e}rale de Lausanne (EPFL), Switzerland, from 2010 to
  2011. From 2011 to 2012 he held a Leverhulme Trust Early Career
  Fellowship at University College London (UCL), where he remains as a
  Newton International Fellow, supported by the Royal Society and the
  British Academy.  His research interests are focused on spherical
  signal processing, including sampling theorems and wavelets on the
  sphere, compressed sensing and Bayesian statistics, and applications
  of these theories to cosmology and radio interferometry.
\end{biographynophoto}

\begin{IEEEbiographynophoto}{Gilles Puy}
  was born in Nevers, France, on August 8, 1985. He received the
  Engineering degree from the Ecole Sup\'erieure d'Electricit\'e
  (Sup\'elec), Gif-sur-Yvette, France, and the M. Sc. degree in
  Electrical and Electronics Engineering from the Ecole Polytechnique
  F\'ed\'erale de Lausanne (EPFL), Lausanne, Switzerland, in 2009.

  In 2009, he joined the Signal Processing Laboratory and the
  Laboratory of functional and metabolic imaging at the EPFL as a
  Doctoral Assistant, where he is currently working towards the
  Ph.D.\ degree in Electrical Engineering. His main research interests
  include inverse problems, compressed sensing, biomedical imaging and
  radio interferometry.
\end{IEEEbiographynophoto}

\begin{IEEEbiographynophoto}{Jean-Philippe Thiran}
  received the Electrical Engineering and Ph.D.\ degrees from the
  Universit\'e catholique de Louvain (UCL), Louvain-la-Neuve, Belgium,
  in 1993 and 1997, respectively.

  Since January 2004, he has been an Assistant Professor, responsible
  for the Image Analysis Group at the Swiss Federal Institute of
  Technology (EPFL), Lausanne, Switzerland. His current scientific
  interests include image segmentation, prior knowledge integration in
  image analysis, partial differential equations and variational
  methods in image analysis, multimodal signal processing, medical
  image analysis, including multimodal image registration,
  segmentation, computer-assisted surgery, and diffusion MRI.

  Dr.\ Thiran was Co-Editor-in-Chief of Signal Processing (published by
  Elsevier Science) from 2001 to 2005. He is currently an Associate
  Editor of the International Journal of Image and Video Processing
  (published by Hindawi), and member of the Editorial Board of Signal,
  Image and Video Processing (published by Springer). He was the
  General Chairman of the 2008 European Signal Processing Conference
  (EUSIPCO 2008). He is a senior member of the IEEE, and a member of
  the MLSP and IVMSP technical committees of the IEEE Signal
  Processing Society.
\end{IEEEbiographynophoto}

\begin{IEEEbiographynophoto}{Pierre Vandergheynst}
  received the M.S. degree in Physics and the Ph.D. degree in
  Mathematical Physics from the Universit\'e catholique de Louvain,
  Louvain-la-Neuve, Belgium, in 1995 and 1998, respectively.

  From 1998 to 2001, he was a Postdoctoral Researcher and an Assistant
  Professor with the Signal Processing Laboratory, Swiss Federal
  Institute of Technology (EPFL), Lausanne, Switzerland. He is now an
  Associate Professor at EPFL, where his research focuses on harmonic
  analysis, sparse approximations, and mathematical image processing
  with applications to higher dimensional, complex data processing.

  Dr. Vandergheynst was co-Editor-in-Chief of Signal Processing from
  2002 to 2006 and has been an Associate Editor of the IEEE
  TRANSACTIONS ON SIGNAL PROCESSING since 2007. He has been on the
  Technical Committee of various conferences and was Co-General
  Chairman of the EUSIPCO 2008 conference. He is the author or
  co-author of more than 50 journal papers, one monograph and several
  book chapters. He is a laureate of the Apple ARTS award and holds
  seven patents.
\end{IEEEbiographynophoto}

\begin{IEEEbiographynophoto}{Dimitri Van De Ville}
  received the M.S. degree in Engineering and Computer Sciences from
  Ghent University, Belgium, in 1998, as well as the Ph.D. degree in
  Computer Science Engineering, in 2002.

  He obtained a grant as Research Assistant with the Fund for
  Scientific Research Flanders Belgium (FWO). In 2002, he joined
  Prof. M. Unser's Biomedical Imaging Group at the Ecole Polytechnique
  F\'ed\'erale de Lausanne (EPFL), Switzerland. In December 2005, he
  became responsible for the Signal Processing Unit at the University
  Hospital of Geneva, Geneva, Switzerland, as part of the Centre
  d'Imagerie Biom\'edicale (CIBM). He was recently awarded a SNSF
  professorship from the Swiss National Science Foundation and he
  currently holds a joint position at the University of Geneva and the
  EPFL. His research interests include wavelets, sparsity, pattern
  recognition, and their applications in biomedical imaging, such as
  functional magnetic resonance imaging.

  Dr. Van De Ville served as an Associate Editor for the IEEE
  Transactions on Image Processing (2006-2009) and the IEEE Signal
  Processing Letters (2004-2006). He is a member of the Bio Imaging
  and Signal Processing (BISP) TC of the IEEE SPS. Since 2003, he has
  also been an Editor and Webmaster of The Wavelet Digest. He is
  co-chair of the Wavelets series conferences (2007, 2009), together
  with V. Goyal and M. Papadakis.
\end{IEEEbiographynophoto}

\begin{biographynophoto}{Yves Wiaux}
  received the M.S. degree in Physics and the Ph.D. degree in
  Theoretical Physics from the Universit\'e catholique de Louvain
  (UCL), Louvain-la-Neuve, Belgium, in 1999 and 2002, respectively.

  He was a Postdoctoral Researcher at the Signal Processing
  Laboratories of the Ecole Polytechnique F\'ed\'erale de Lausanne
  (EPFL), Switzerland, from 2003 to 2008. He was also a Postdoctoral
  Researcher of the Belgian National Science Foundation (F.R.S.-FNRS)
  at the Physics Department of UCL from 2005 to 2009. He is now a
  Ma\^itre Assistant of the University of Geneva (UniGE), Switzerland,
  with joint affiliation between the Institute of Electrical
  Engineering and the Institute of Bioengineering of EPFL, and the
  Department of Radiology and Medical Informatics of UniGE. His
  research lies at the intersection between complex data processing
  (including development on wavelets and compressed sensing) and
  applications in astrophysics (notably in cosmology and radio
  astronomy) and in biomedical sciences (notably in MRI and diffusion
  MRI).
\end{biographynophoto}

% You can push biographies down or up by placing
% a \vfill before or after them. The appropriate
% use of \vfill depends on what kind of text is
% on the last page and whether or not the columns
% are being equalized.

%\vfill

% Can be used to pull up biographies so that the bottom of the last one
% is flush with the other column.
%\enlargethispage{-5in}

\end{document}